\documentclass[a4paper,11pt]{article}

\usepackage{amssymb,amsmath}
\usepackage[dvips]{graphicx}
\usepackage{macro,cite}
\usepackage{arydshln}

\renewcommand{\Proof}{{\it Proof\,:}\hspace{0.2cm}}

\newcommand{\one}{\mathbf{1}}

\begin{document}

\title{\LARGE
A Web Aggregation Approach for \\
Distributed Randomized PageRank Algorithms\thanks{%
This paper has been accepted for publication in the
IEEE Transactions on Automatic Control, 2012.
This work was supported
in part by the Ministry of Education, Culture, Sports, Science and Technology
in Japan under Grant-in-Aid for Scientific Research, No.~21760323, 
the European Union Seventh Framework Programme [FP7/2007-2013]  under
grant agreement n$^\circ$257462 HYCON2 Network of Excellence, and PRIN 2008 of
Ministry of Education, Universities and Research of Italy.}}

\author{%
Hideaki Ishii\\[1mm]
Department of Computational Intelligence 
and Systems Science\\
Tokyo Institute of Technology\\
4259 Nagatsuta-cho, Midori-ku, Yokohama 226-8502, Japan\\
E-mail: ishii@dis.titech.ac.jp\\[4mm]
Roberto Tempo\\[1mm]
CNR-IEIIT, Politecnico di Torino\\
Corso Duca degli Abruzzi 24, 10129 Torino, Italy\\
E-mail: roberto.tempo@polito.it\\[4mm]
Er-Wei Bai\\[1mm]
Department of Electrical and Computer Engineering, The University of Iowa\\
4316 Seamans Center for the Engineering Arts and Sciences\\
Iowa City, IA 52242-1527, U.S.A.\\
and School of Electronics, Electrical Engineering and Computer Science\\
Queen's University, Belfast, BT7 1NN, U.K.\\
E-mail: er-wei-bai@uiowa.edu}

\addtolength{\abovedisplayskip}{-.3mm}
\addtolength{\belowdisplayskip}{-.3mm}
\addtolength{\jot}{-.2mm}
\addtolength{\textfloatsep}{-2mm}
\addtolength{\floatsep}{2mm}
\addtolength{\parsep}{12mm}
\setlength{\baselineskip}{3.4ex}

\maketitle


\begin{abstract}
The PageRank algorithm employed at Google assigns 
a measure of importance to each web page for rankings
in search results. In our recent papers, we have proposed a
distributed randomized approach for this algorithm, where
web pages are treated as agents computing their own 
PageRank by communicating with linked pages. 
This paper builds upon this approach 
to reduce the computation and communication loads
for the algorithms.
In particular, we develop a method to 
systematically aggregate the web pages into groups by
exploiting the sparsity inherent in the web.
For each group, an aggregated PageRank value is computed,
which can then be distributed among the group members.
We provide a distributed update scheme for the aggregated
PageRank along with an analysis on its convergence properties.
The method is especially motivated by 
results on singular perturbation techniques for large-scale 
Markov chains and multi-agent consensus. 
%
\end{abstract}


\section{Introduction}
\label{sec:intro}

When using the search engine Google, the rankings in search results 
take account of various aspects of web pages, but it has been
acknowledged that the so-called PageRank algorithm provides
crucial information for this purpose. 
This algorithm assigns to each web page 
a measure of its importance or popularity based
solely on the link structure of the web. In particular,
pages possessing more links, especially those from important pages, 
are given higher PageRank values, increasing the chance to
be placed on the top of search results 
(see, e.g., \cite{BriPag:98,BryLei:06,LanMey:06}).

One of the main challenges in implementing this algorithm is the size 
of the web. It is reported that the number of web page indices 
collected at Google is over 10 billion, causing serious 
issues for computation. 
Numerical methods for PageRank have been a subject of recent research.
In the adaptive scheme of \cite{KamHavGol:04}, 
computational resources are allocated to pages 
whose convergence to the PageRank values is slow.
In \cite{LinShiWei:09}, the problem size is reduced by 
treating the set of the so-called dangling nodes as a single node.
The work of \cite{AvrLitNem:07} employs techniques based on 
Monte Carlo simulation. On the other hand, 
numerical analysis methods known as asynchronous iterations 
\cite{BerTsi:89} are applied to PageRank algorithms in 
\cite{deJBra:07,KolGalSzy:06}.
In \cite{NazPol:11}, a randomized algorithm is proposed
based on stochastic descent methods with an explicit bound 
on the convergence rate. 
Moreover, the variations in PageRank values when the link structure
changes have been studied 
from the viewpoint of fragile/uncertain links in \cite{IshTem_sice:09}
and also from optimization and linear algebra viewpoints 
in \cite{CsaJunBlo:09,deKNinVan:08}\;\footnote{%
In the area of e-commerce, various methods are known to enhance 
chances of specific web pages to be placed higher in search results, 
e.g., by including effective keywords in the pages. 
Such methods are sometimes
referred to as search engine optimization (e.g., \cite{ChaLakFri:09}).
In these methods, the importance of PageRank is often emphasized 
through adding proper links.}.

In our recent paper \cite{IshTem:10}, we focused 
on this algorithm and developed a distributed randomized approach 
for PageRank computation. 
From the control theoretic viewpoint, a key observation 
is that the PageRank computation shares several features
with the multi-agent consensus problems, 
which have gained much attention in recent years; 
see, e.g., \cite{csm:07,AntBai:07,BerTsi:89} and the references therein.
Thus, we view the web as a network of agents having computation 
and communication capabilities. The idea is to let each web page,
or the server that hosts it, compute its own PageRank value 
by communicating with neighboring pages connected by direct links.  
To realize asynchronous communication, 
it employs the so-called gossip protocol, where
the pages randomly determine when 
information should be transmitted.
Such a randomization-based method is motivated by the recent 
advances on probabilistic methods in systems and control \cite{TemCalDab_book}
and has been adopted in the literature of multi-agent consensus 
(e.g., \cite{BoyGhoPra:06,CaiIsh:11,CarFagFoc:08,FagZam:08,
HatMes:05,KasBasSri:07,TahJad:08,TemIsh:07,Wu:06}).
In \cite{IshTem_yyfest:10,IshTemBai:10}, we have also considered the effects of 
communication failures among the pages under this approach.

This paper aims at generalizing 
the distributed PageRank algorithms in \cite{IshTem:10}
so that they can be used in an environment under limited resources.
In particular, we develop efficient algorithms by reducing the
amount of computation and communication loads.
In such an environment, the computation of the true PageRank values
may be difficult. Consequently, we provide an alternative method
for finding a good approximate with bounds on the possible errors.

The proposed approach is based on a novel aggregation method
of the original web to reduce the size of the problem.
The pages are first divided into a number of groups, for example,
based on the hosts or the domains of the pages. 
It is known that most of the links in the web are 
intra-host ones \cite{BroLem_infret:06,LanMey:06},
and thus the underlying graph has certain sparsity properties.
To exploit such properties, we further aggregate the graph
so that each group either (i)~has more internal links than those going 
outside or (ii)~consists of just one page. The aggregation procedure is 
easy to implement, employing a simple criterion, and can be applied to 
graphs with any link structures.
Then, each group computes only one value in a decentralized manner
via an enhanced version of the algorithms in \cite{IshTem:10}.
This value represents the total value of the group members.
Once this is computed, it can be distributed among the group members 
to determine their individual values.
It is demonstrated theoretically and also 
through a numerical example that aggregation 
can reduce the computational cost while maintaining 
the accuracy and the convergence rate at a level similar to 
the non-aggregated full-order case. 

The aggregation technique is particularly motivated by the
singular perturbation analyses for large-scale systems
in Markov chains \cite{AldKha:91,PhiKok:81} 
(see also \cite{CodWilSasCas_stoch:83,YinZha:98})
and multi-agent consensus type problems \cite{BiyArc:08,ChoKok:85}.
It is important to note that common in these works is the strong
assumption on interactions among groups, requiring
\textit{all groups} to have only limited ratios of 
outgoing links towards other groups.
This would not hold for the web graph where
pages with many external links may not be common but indeed do exist.
In the proposed aggregation method, such pages are treated as exceptions
and are separated into groups of their own.
We will later discuss the relation of our approach to those using
singular perturbation techniques in more detail. 
Aggregation for PageRank computation has also been explored 
to obtain acceptable approximation
in \cite{LanMey:06,ZhuYeLi:05} by classical methods 
in the Markov chain literature.
The paper \cite{BroLem_infret:06} examines aggregation 
through extensive simulation using large data of the web.
Recent works on aggregation for Markov chains can be found, 
e.g., in \cite{DenMehMey:09}.
More generally, in the literature of complex networks, 
partitioning graphs into communities is a topic 
widely studied under various criteria for detecting communities; 
see, for example, \cite{DelYalBar:10} and the references therein.

From the viewpoint of distributed randomized approach, 
there are two new features in the current paper. 
First, the nodes initiate updates in a random manner
as in the original algorithm of \cite{IshTem:10}.
The difference is that an updating page transmits its value 
only to the pages to which it has outgoing links;
this means that the extra data required in the previous algorithms
on pages having incoming links towards this page becomes unnecessary.
Second, each node can further divide its linked pages
into several groups and communicates with them separately
under possibly different update probabilities.
These features are useful to reduce the amount of the overall communication,
especially for pages with many links.


This paper is organized as follows: 
We first give a brief overview of the PageRank problem 
in Section~\ref{sec:pagerank}. 
In Section~\ref{sec:prob}, we introduce the approach for 
web aggregation and the communication protocol among agents
and then formulate the problem of
computing the PageRank values via web aggregation.
In Section~\ref{sec:agg}, the main results on the 
aggregation-based algorithm are presented along with
an analysis on error bounds.
Discussions on the relation to singular perturbation techniques
are given in Section~\ref{sec:sing}.
In Section~\ref{sec:distributed}, the distributed randomized 
approach is developed for the part of the proposed algorithm 
based on the reduced-order recursion.
We provide a numerical example in Section~\ref{sec:example} to
illustrate the results. 
The paper is finally concluded in Section~\ref{sec:concl}.
Parts of the results in this paper have appeared in 
preliminary forms in \cite{IshTemBai:10,IshTemBaiDab:09}.

\smallskip
\noindent
{\it Notation}:~~For vectors and matrices, inequalities 
are used to denote entry-wise inequalities:
For $X,Y\in\R^{n\times m}$, $X\leq Y$ implies
$x_{ij}\leq y_{ij}$ for $i=1,\ldots,n$ and $j=1,\ldots,m$;
in particular, we say that the matrix $X$ is nonnegative if $X\geq 0$ 
and positive if $X> 0$.
A probability vector is a nonnegative vector $v\in\R^n$ such that
$\sum_{i=1}^n v_i = 1$.
A matrix $X\in\R^{n\times n}$ is said to be (column) stochastic
if it is nonnegative and each column sum equals 1, i.e.,
$\sum_{i=1}^n x_{ij}=1$ for each $j$.
Let $\one_n\in\R^n$ be the vector whose entries are all $1$ as
$\one_n:=[1\,\cdots\,1]^T$.
Similarly, $S\in\R^{n\times n}$ is the matrix with all 
entries being $1$.
The spectral radius of the matrix $X\in\R^{n\times n}$ is
denoted by $\rho(X)$.  

\section{The PageRank problem}
\label{sec:pagerank}

In this section, the PageRank problem is briefly described
based on, e.g., \cite{BriPag:98,BryLei:06,LanMey:06}.

Consider the directed graph $\Gcal=(\Vcal,\Ecal)$ representing
a network of $n$ web pages.
Here, $\Vcal:=\{1,2,\ldots,n\}$ is the set of nodes corresponding 
to the web page indices
while $\Ecal\subset\Vcal\times\Vcal$ is the set of edges 
for the links among the pages. 
The node $i$ is connected to the node $j$ by an edge, i.e.,
$(i,j)\in\Ecal$, if page $i$ has an outgoing link to page $j$.

The objective of the PageRank algorithm is to assign some
measure of importance to each web page.  
The Page\-Rank value of page $i\in\Vcal$
is given by $x_i^*\in[0,1]$.
The relation $x_i^*>x_j^*$ implies that page $i$ has
higher rank than page $j$.
The pages are ranked according to the rule that
a page having more links, especially those from important pages, 
becomes more important.
This is done in such a way that the value of one page equals the 
sum of the contributions from all pages that have links to it.  
Let the values be in the vector form as $x^*\in[0,1]^n$.
Then, the PageRank vector $x^*$ is defined by
\begin{equation}
   x^* = A x^*,~~x^*\in[0,1]^n,~~\one_n^T x^* = 1,
\label{eqn:xA:pr}
\end{equation}
where the link matrix $A=(a_{ij})\in\R^{n\times n}$ is given by
$a_{ij}=1/n_j$ if $(j,i)\in\Ecal$ and 0 otherwise,
and $n_j$ is the number of outgoing links of page $j$.
Hence, the value vector $x^*$ is a nonnegative unit 
eigenvector corresponding to the eigenvalue 1 of $A$.

In general, for this eigenvector to exist and
then to be unique, it is sufficient that 
the web as a graph is strongly connected \cite{HorJoh:85}
\footnote{A directed graph is said to be strongly connected
if for any two nodes $i,j\in\Vcal$, there exists a sequence of
edges which connects node $i$ to node $j$.}.
However, the web is known not to be strongly connected.
Thus, the convention is to slightly modify the problem as follows.
First, to simplify the discussion, we redefine the graph, and thus the
matrix $A$, by bringing in artificial links for 
nodes with no outgoing links such as PDF files.
This can be done by adding links back to the pages having links
to such pages.
As a result, the link matrix $A$ becomes a stochastic matrix,
that is, $\sum_{i=1}^n a_{ij}=1$ for each $j$.
This implies that there exists at least one eigenvalue equal to 1. 
To guarantee the uniqueness of this eigenvalue,
let $m$ be a parameter such that $m\in(0,1)$, 
and let the modified link matrix $M\in\R^{n\times n}$ be defined by
$M := (1-m)A + (m/n)S$, where $S\in\R^{n\times n}$ is the matrix whose 
entries are all 1.
Notice that $M$ is a positive stochastic matrix%
\footnote{In the original algorithm in \cite{BriPag:98}, 
a typical value for $m$ is reported to be $m=0.15$,
but no specific reason is given for this choice.
We will use this value throughout this paper.}.
By Perron's theorem \cite{HorJoh:85}, 
the eigenvalue 1 is of multiplicity 1 and 
is the unique eigenvalue with maximum magnitude.
Further, the corresponding eigenvector is positive. 
Hence, we redefine the value vector $x^*$ by using $M$ 
as follows.

\begin{definition}\rm\label{def:pagerank}
The PageRank value vector $x^*$ is given by 
\begin{equation}
   x^* = M x^*,~~x^*\in[0,1]^n,~~\one_n^T x^* = 1.
 \label{eqn:prvec}
\end{equation}
\end{definition}

Due to the large dimension of the link matrix $M$, the computation
of $x^*$ is difficult. 
The solution employed in practice is based on the power method 
given by the recursion 
\begin{align}
  x(k+1) 
    &= M x(k) 
     = (1-m)A x(k) + \frac{m}{n}\one_n,
\label{eqn:xM0}
\end{align}
where $x(k)\in\R^n$ and 
the initial vector $x(0)\in\R^n$ is a probability vector. 
The second equality above follows from the fact 
$Sx(k)=\one_n$, $k\in\Z_+$.
For implementation, the form on the far right-hand side
is important, using only the sparse matrix $A$ and not the dense 
matrix $M$. This method asymptotically finds 
the value vector as shown below \cite{HorJoh:85}.

\begin{lemma}\rm\label{lem:power1}
In the update scheme \eqref{eqn:xM0},
for any initial state $x(0)$ that is a probability vector, 
it holds that $x(k)\rightarrow x^*$ as $k\rightarrow\infty$.
\end{lemma}


\section{Problem formulation}
\label{sec:prob}

In this section, we introduce the problem setting
for the distributed computation of the aggregated PageRank.
Following the randomized distributed approach 
proposed in \cite{IshTem:10}, we view the web as a
network of agents having computation and communication
capabilities. 
The focus here is to extend the distributed algorithm of
\cite{IshTem:10}
so that it can be executed with reduced computation and communication
to determine approximate values of the exact PageRank.
In what follows, we present the procedure for aggregation
of the web and then introduce the communication protocol.

\subsection{Web aggregation}
\label{sec:prob:agg}

In the proposed approach, 
the original web is aggregated by assigning each page into
a number of groups and then
each group computes one value,
which is the sum of the values of the group members.
We aggregate the pages sharing the following three properties:
(i)~The pages are placed under the same host/server so that their
values can be computed together.
(ii)~Each group has a sufficiently large number of 
internal links. More specifically,
pages have more links within their own groups
than those pointing at pages that 
belong to other groups having multiple members. 
(iii)~Group members are expected to take similar 
values in PageRank; this may be known from
past computations and/or the link structure.
We will show that the process of grouping can be done locally 
at each host.

We develop a novel aggregation approach by exploiting sparsity 
properties that the web inherently has, as stated by (ii) above.
The particular approach has a close relation to the singular perturbation 
analysis for large-scale systems with network structures
\cite{AldKha:91,BiyArc:08,ChoKok:85,PhiKok:81}.
The method there however requires a stronger sparsity notion 
on the underlying graph, which seems difficult to expect in the web. 
Hence, necessary modifications will be made in the approach.
The relation among these papers will be discussed in Section~\ref{sec:sing}.

First, partition the original web graph $\Gcal=(\Vcal,\Ecal)$ 
and construct the \textit{aggregated} graph denoted by 
$\widetilde{\Gcal}=(\widetilde{\Vcal},\widetilde{\Ecal})$ 
as follows:
\begin{enumerate}
\item[(i)] 
The node set is given by $\widetilde{\Vcal}:=\{1,2,\ldots,r\}$, 
and each node $i$ represents a partition set $\Ucal_i$ of $\Vcal$, that is,
$\bigcup_i\Ucal_i = \Vcal$ and 
$\Ucal_i\cap\Ucal_j=\emptyset$, $\forall i\neq j$.
We call the set $\Ucal_i$ a group of pages.
Let $r$ be the number of groups, and 
let $\widetilde{n}_i$ be the number of pages in group $\Ucal_i$.
Thus, $\sum_{i=1}^{r}\widetilde{n}_i = n$.
\item[(ii)] The edge set $\widetilde{\Ecal}=\widetilde{\Vcal}\times\widetilde{\Vcal}$
satisfies that if $(i_1,i_2)\in\Ecal$, then 
$(h(i_1),h(i_2))\in\widetilde{\Ecal}$, where
$h:\Vcal\rightarrow\widetilde{\Vcal}$ is the function indicating 
the group $j$ that the web page $i$ belongs to such that 
$h(i)=j$, or $i\in\Ucal_j$.
\end{enumerate}
To simplify the notation, without loss of generality, 
we assume that in the PageRank vector $x^*$, the first $\widetilde{n}_1$ 
entries correspond to the pages belonging to group $\Ucal_1$,
and the following $\widetilde{n}_2$ entries are for those 
in group $\Ucal_2$, and so on.

We make the following assumption regarding the grouping.
It says that each group should have a sufficiently small number 
of external links compared to internal ones.
Recall from \eqref{eqn:xA:pr} that $n_i$ denotes the number of 
outgoing links of page $i$, 
and let $n_{\text{ext},i}$ be the number of outgoing links from 
page $i$ to groups having more than one page. Following
\cite{ChoKok:85}, we define the \textit{node parameter} $\delta_i$
of page $i$ by
\begin{equation}
  \delta_i 
    := \frac{n_{\text{ext},i}}{n_i},~~i=1,\ldots,n.
  \label{eqn:delta_i}
\end{equation}

\begin{assumption}\label{assump:delta}\rm
Given the bound $\delta\in(0,1)$ on node parameters, 
each group $j$ satisfies one of the following conditions:
\begin{enumerate}
 \item[(i)] 
       For each page $i$ in group $j$, 
       it holds that $\delta_i \leq \delta$.
 \item[(ii)]        
       Group $j$ consists of only one page.
\end{enumerate}  
\end{assumption}
In view of (ii) above, groups with one member are called \textit{single}
groups; denote by $r_1$ the number of such groups. These groups
represent exceptional pages having high ratios of external links.

After the groups of pages are determined satisfying the
assumptions above, we consider the values that represent the groups.
For this purpose, in the update scheme, 
we employ the coordinate transformation
$\widetilde{x}(k):=Vx(k)$
via the matrix $V = \big[V_1^T~V_2^T\big]^T\in\R^{n\times n}$ given by 
\begin{align}
\begin{split}
 V_1 &:= \text{bdiag}
            (\one_{\widetilde{n}_i}^T)
            \in\R^{r\times n},\\
 V_2 &:= \text{bdiag}
         \Big(
          [I_{\widetilde{n}_i-1}~0]
            - \frac{1}{\widetilde{n}_i} 
               \one_{\widetilde{n}_i-1}\one_{\widetilde{n}_i}^T
         \Big)
         \in\R^{(n-r)\times n},
\end{split}
\label{eqn:V}
\end{align}
where $\text{bdiag}(X_i)$ denotes a block-diagonal matrix
whose $i$th diagonal block is $X_i$.
It should be noted that $V_1$ and $V_2$ are block-diagonal matrices 
containing $r$ and $r-r_1$ blocks, respectively. 
They have simple structures, depending only on the sizes $\widetilde{n}_i$ 
of the groups. 
Note that in $V_2$, if the $i$th group is a single one
(i.e., $\widetilde{n}_i=1$), then the $i$th block has the size
$0\times 1$, meaning that the corresponding column is zero. 
Moreover, $V_1$ and $V_2$ are orthogonal: $V_1 V_2^T = 0$.

The PageRank vector $\widetilde{x}^*$ and 
the state $\widetilde{x}(k)$ after the transformation are
partitioned as
\begin{equation}
  \widetilde{x}^*
   = \begin{bmatrix}
        \widetilde{x}_1^*\\
        \widetilde{x}_2^*
     \end{bmatrix}
   := \begin{bmatrix}
         V_1\\ V_2
      \end{bmatrix} x^*,~~~
  \widetilde{x}(k)      
  = \begin{bmatrix}
      \widetilde{x}_1(k)\\
      \widetilde{x}_2(k)
    \end{bmatrix}
  := \begin{bmatrix}
        V_1 \\
        V_2
     \end{bmatrix} x(k).
\label{eqn:xtilde}
\end{equation}
In the first part $\widetilde{x}_1^*$ of
the PageRank vector in the new coordinate, the $i$th entry is 
the total value of the members in group $i$;
this vector $\widetilde{x}_1^*$ is referred to as the
\textit{aggregated PageRank}.
In the second part $\widetilde{x}_2^*$, 
each entry represents the difference between 
a page value and the average value of the group members.
In the distributed algorithm developed in Section~\ref{sec:distributed},
the objective is to compute the aggregated PageRank $\widetilde{x}_1^*$ 
in a recursive fashion
via information exchange only among groups.
After this is completed, the second part $\widetilde{x}_2^*$ 
should be obtained. It will be shown that in this stage, 
transmissions among pages in different
groups is necessary, but only once during the execution of the algorithm.
Hence, reduced communication load can be expected 
in particular when $r$ is small.

\begin{remark}\label{rem:group}\rm
For a given bound $\delta$ on the node parameters, 
a simple grouping procedure for
Assumption~\ref{assump:delta} to hold can be described as follows. 
The pages are initially grouped based on their hosts, so 
the computation of the node parameters $\delta_i$ in \eqref{eqn:delta_i}
can be done locally. Any page $i$ whose $\delta_i$ does not
satisfy the condition (i) is taken out from the group;
such pages are treated as single groups, for which the condition 
(ii) applies. Other pages
still belong to the same group, and thus their parameters $\delta_i$ 
are updated and then checked whether (i) holds for this new group.
These steps are repeated until all pages under one host satisfy
the assumption. It is clear that 
for any given bound $\delta$ on the node parameters,
this procedure will terminate.
It should be noted that there is a tradeoff between 
the parameter $\delta$ and the number $r$ of groups:
Smaller $\delta$ implies larger $r$, and vice versa.
\End
\end{remark}

\subsection{Communication protocol via random gossipping}
\label{sec:prob:channel}

We next discuss the communication among groups
in the proposed distributed algorithm.

For the computation of $\widetilde{x}_1(k)$,
the groups send their values to linked groups.
Here, we employ a gossip-type protocol,
where the groups
decide to communicate with their linked neighbors at random
times. Such a protocol is based on local information only
and does not require a common clock.
It is thus useful in realizing asynchronous 
algorithms for a network of agents
(e.g., \cite{BoyGhoPra:06,CaiIsh:11,CarFagFoc:08,FagZam:08,
HatMes:05,KasBasSri:07,TahJad:08,TemIsh:07,Wu:06}).




In the aggregated graph $\widetilde{\Gcal}=(\widetilde{\Vcal},\widetilde{\Ecal})$,
the nodes exchange their values over their outgoing links.
Denote by $\widetilde{\Vcal}_i$ the set of indices of the 
groups having links from node $i$ as
\[
  \widetilde{\Vcal}_i 
   := \bigl\{
        j\in\widetilde{\Vcal}:~(i,j)\in\widetilde{\Ecal},~j\neq i
      \bigr\}.
\]
Here, we allow node $i$ to communicate with a subset of 
$\widetilde{\Vcal}_i$ at a time. This helps to reduce 
the instantaneous communication load especially for nodes 
having many links.
For this purpose, we partition $\widetilde{\Vcal}_i$ into the sets
$\widetilde{\Vcal}_{i,1},\ldots,\widetilde{\Vcal}_{i,g_i}$,
where $g_i$ is the number of partition sets, i.e., it holds that 
\begin{equation*}
  \widetilde{\Vcal}_i 
      = \bigcup_{\ell=1}^{g_i} \widetilde{\Vcal}_{i,\ell},~~~~
  \widetilde{\Vcal}_{i,\ell}\cap \widetilde{\Vcal}_{i,j} 
      = \emptyset,~~\forall\ell\neq j.
\end{equation*}
For each node $i\in\widetilde{\Vcal}$, 
let $\eta_i(k)\in\{0,1,\ldots,g_i\}$ be
the i.i.d.\ random process 
that specifies the set of nodes to which it sends the value 
$(\widetilde{x}_1(k))_i$
at time $k$. That is,
\begin{equation}
  \eta_i(k) 
    = \begin{cases}
        \ell & \text{if node $i$ sends its value to nodes 
               in $\widetilde{\Vcal}_{i,\ell}$, $\ell=1,2,\ldots,g_i$},\\
        0    & \text{if node $i$ does not communicate}
      \end{cases}
      \label{eqn:eta}
\end{equation}
for $k\in\Z_+$.  The probability distribution of this process is given as
\begin{equation}
  \alpha_{i,\ell}
    = \Prob \{
        \eta_i(k) = \ell
         \},~~~\ell = 0,1,\ldots,g_i,~~k\in\Z_+.
 \label{eqn:alpha_agg}
\end{equation}
The update probabilities $\alpha_{i,\ell}\in(0,1)$ 
are chosen so as to satisfy the condition
\begin{equation}
  \sum_{\ell=0}^{g_i} \alpha_{i,\ell}  
     = 1,~~~i\in\widetilde{\Vcal}.
\label{eqn:alpha_il}
\end{equation}

\medskip
The main problem studied in this paper can be roughly restated 
as follows:
Design a distributed randomized algorithm for computing
approximated PageRank values such that 
(i)~the groups compute $\widetilde{x}_1(k)$, 
the total values of their member pages,
following the gossip protocol for communication 
and then 
(ii)~from $\widetilde{x}_1(k)$, the PageRank vector $x(k)$
and, in particular, the values for individual
pages are obtained. 

We characterize the web aggregation approach in 
Section~\ref{sec:agg} along with error analyses for the
aggregated PageRank. Then, in Section~\ref{sec:distributed},
the distributed randomized algorithm of reduced order
for computing the group values $\widetilde{x}_1(k)$ is discussed.

\section{Aggregation-based PageRank computation}
\label{sec:agg}

In this section, we present the approach for aggregating
the web graph and then propose 
an approximated version of the PageRank that can
be computed from a lower-order update scheme.

\subsection{Definition of aggregated PageRank}
\label{sec:agg:def}

We begin by analyzing the centralized update scheme 
of \eqref{eqn:xM0} described in Section~\ref{sec:pagerank} 
when the state is transformed as
$\widetilde{x}(k)=Vx(k)$ by \eqref{eqn:xtilde}.
Let $\widetilde{A}:=VAV^{-1}$ be the link matrix in the 
new coordinate. Partition it 
in accordance with the dimensions of $\widetilde{x}_1(k)$ and 
$\widetilde{x}_2(k)$ as
\begin{align}
  \widetilde{A}
    &= \begin{bmatrix}
          \widetilde{A}_{11} & \widetilde{A}_{12}\\
          \widetilde{A}_{21} & \widetilde{A}_{22}
       \end{bmatrix}
  \label{eqn:Atil0}
\end{align}
with $\widetilde{A}_{11}\in\R^{r\times r}$.
The update scheme in \eqref{eqn:xM0} can be expressed as
\begin{align}
  \widetilde{x}_1(k+1)
   &= (1-m) \widetilde{A}_{11} \widetilde{x}_1(k)
       + (1-m) \widetilde{A}_{12} \widetilde{x}_2(k) + \frac{m}{n}u,
   \label{eqn:updateVx1}\\
  \widetilde{x}_2(k+1)
   &= (1-m) \widetilde{A}_{21} \widetilde{x}_1(k)
        + (1-m) \widetilde{A}_{22} \widetilde{x}_2(k),
   \label{eqn:updateVx2}
\end{align}
where $u:=V_1\one_n=[\widetilde{n}_1\;\cdots\widetilde{n}_r]^T$; 
we also used the fact $V_2\one_n = 0$.
The initial states are such that $\widetilde{x}_1(0)\geq 0$ and
$\one_r^T\widetilde{x}_1(0)=1$.
The steady state of this scheme 
is the transformed PageRank vector
$\widetilde{x}^*$ given in \eqref{eqn:xtilde}.

Now, to derive an approximated version of the update scheme above, 
we focus on the characteristics of the submatrices $\widetilde{A}_{ij}$.
The transformation matrix $V$ in \eqref{eqn:V} has a simple 
structure, and the advantage is that 
its inverse can be found in an explicit form, 
which will be useful in our analysis.
Denote the inverse by $W:=V^{-1}$. It can be partitioned as
$W = \big[W_1~W_2\big]$ where
\begin{align*}  
  W_1 
   &:= \text{bdiag}
         \Big(
           \frac{1}{\widetilde{n}_i}\one_{\widetilde{n}_i}
         \Big)
          \in\R^{n\times r},\\
  W_2 
   &:= \text{bdiag}
         \Bigg(
           \begin{bmatrix}
             I_{\widetilde{n}_i-1}\\
             -\one_{\widetilde{n}_i-1}^T
           \end{bmatrix}
         \Bigg)
           \in\R^{n\times (n-r)}.
\end{align*}
Again, $W_1$ and $W_2$ are block-diagonal matrices with
$r$ and $r-r_1$ blocks, respectively. 
Moreover, the rows in $W_2$ that correspond to single groups
are zero. It is obvious that $V_1W_1=I$, $V_1W_2=0$, $V_2W_1=0$, and
$V_2W_2=I$. 

Based on the approach studied in \cite{PhiKok:81},
the key observation in the proposed aggregation is that
the matrix $A$ can be decomposed into three parts as 
\begin{equation}
  A = I + A_{\text{int}} + A_{\text{ext}}.
  \label{eqn:Adecomp}
\end{equation}
Here, the \textit{internal} link matrix $A_{\text{int}}$ is block diagonal; 
its $i$th block
is of the size $\widetilde{n}_i\times \widetilde{n}_i$, whose nondiagonal
entries are the same as those of $A$, but its diagonal entries 
are chosen so that the column sums are zero.
This implies that $I+A_{\text{int}}$ is a block-diagonal stochastic matrix. 
Hence, it easily follows that 
\begin{equation}
  V_1 A_{\text{int}} = 0.
  \label{eqn:V1AI}
\end{equation}
On the other hand, the \textit{external} link matrix $A_{\text{ext}}$ 
contains all elements in $A$
which are not in the block-diagonal $A_{\text{int}}$ while its
diagonal entries are chosen so that each column sum equals
zero. Let $A_{\text{ext0}}$ be an $n\times n$ matrix whose
$j$th column is the same as that of $A_{\text{ext}}$ if page $j$
belongs to a non-single group (i.e., with more than one member) 
and zero otherwise
for $j=1,\ldots,n$.
By the definition of $W_2$, it is simple to check that 
\begin{equation}
   A_{\text{ext}} W_2=A_{\text{ext0}} W_2.
\label{eqn:AE}
\end{equation}

By using the facts $W=V^{-1}$, \eqref{eqn:Adecomp}, \eqref{eqn:V1AI},
and \eqref{eqn:AE},
the submatrices of $\widetilde{A}$ in \eqref{eqn:Atil0}
can be expressed as
\begin{align}
  \begin{bmatrix}
    \widetilde{A}_{11} & \widetilde{A}_{12}\\
    \widetilde{A}_{21} & \widetilde{A}_{22}
  \end{bmatrix}
    = \begin{bmatrix}
          V_1 A W_1 & V_1 A W_2\\
          V_2 A W_1 & V_2 A W_2
       \end{bmatrix}
    = \begin{bmatrix}
        I + V_1 A_{\text{ext}} W_1   & V_1 A_{\text{ext0}} W_2\\
        V_2 (A_{\text{int}}+A_{\text{ext}}) W_1 & I + V_2 (A_{\text{int}} + A_{\text{ext0}}) W_2
      \end{bmatrix}.
\label{eqn:Atil}
\end{align}
For later use, from $\widetilde{A}_{22}$, 
we remove $A_{\text{ext0}}$ to obtain
the block-diagonal matrix $\widetilde{A}'_{22}$ given by
\begin{equation}
  \widetilde{A}'_{22} 
   := I + V_2 A_{\text{int}} W_2.
\label{eqn:A22dash}
\end{equation}

The following results are helpful to justify the approach, as we shall see later.

\begin{lemma}\label{lem:AE}\rm
 \begin{enumerate}
 \item[(i)] The matrix $\widetilde{A}_{11}$ is stochastic. 
 \item[(ii)] 
             The matrix $\widetilde{A}'_{22}$ in \eqref{eqn:A22dash}
             has spectral radius smaller than or equal to 1.
 \item[(iii)] Under Assumption~\ref{assump:delta}, it holds that
             $\norm{A_{\text{ext0}}}_1 \leq 2\delta$. 
\end{enumerate} 
\end{lemma}

\Proof
(i)~It is clear that $\widetilde{A}_{11}$ is nonnegative and satisfies
$\one_r^T \widetilde{A}_{11} = \one_r^T V_1 A W_1 =\one_r^T$. 


(ii)~The matrix $I+A_{\text{int}}$ is stochastic. In particular, it has
$r$ diagonal blocks, which are all stochastic. Thus, $I+A_{\text{int}}$ has
at least $r$ eigenvalues equal to 1, and the rest have magnitude 
less than or equal to 1.
The transformation matrix $V$ is composed of orthogonal rows,
and thus $\Image V_2^T = (\Image V_1^T)^{\perp}$. However, 
$\Image V_1^T$ is the left eigenspace of $I+A_{\text{int}}$ corresponding 
to $r$ of the eigenvalues 1.
This implies that $\Image V_2^T$ spans the eigenspace for 
the rest of the eigenvalues of $I+A_{\text{int}}$;
let $\lambda$ be any such eigenvalue. Then, there exists
a vector $v\in\C^{n-r}$ such that $v^T V_2 (I+A_{\text{int}})=\lambda v^T V_2$.
Notice $V_2 W_2 = I$, and thus
$v^T V_2 (I+A_{\text{int}})W_2=\lambda v^T$, showing that
$\lambda$ is an eigenvalue of $V_2 (I+A_{\text{int}}) W_2$ as well.
Therefore, we conclude that 
$\rho(\widetilde{A}'_{22})=\rho(V_2 (I+A_{\text{int}}) W_2)\leq 1$. 

(iii)~For $i\neq j$, the $(i,j)$ entry of $A_{\text{ext0}}$ 
is nonzero (and equals $a_{ij}=1/n_j$) if and only if page $j$ has 
a link to page $i$, and moreover pages $i$ and $j$ belong to different 
groups, each of which having multiple group members. 
By assumption, for each column, the sum of its off-diagonal entries is 
less than or equal to $\delta$, but each column sum equals zero.
Hence, the 1-norm of $A_{\text{ext0}}$ is bounded by $2\delta$.
\EndProof

\medskip
An important implication of (ii) and (iii) of this lemma is that 
if the node parameter $\delta$ in Assumption~\ref{assump:delta}
is sufficiently small,
then the matrix $(1-m)\widetilde{A}_{22}$ is stable%
\footnote{A matrix is said to be stable if it is Schur stable, 
that is, if all of its eigenvalues have magnitude less than 1.};
this is because from \eqref{eqn:Atil}, we have
$\widetilde{A}_{22}=I+V_2(A_{\text{int}}+A_{\text{ext0}})W_2
=\widetilde{A}_{22}'+V_2 A_{\text{ext0}} W_2$, where $A_{\text{ext0}}$
is proportional to $\delta$; recall also that $\widetilde{A}_{22}'$
is a block-diagonal matrix, which will become crucial from the
computational viewpoint.
We now come to the idea of how to approximate the scheme
\eqref{eqn:updateVx1} and \eqref{eqn:updateVx2}.
First, express \eqref{eqn:updateVx2} for $\widetilde{x}_2(k)$ 
using its steady state 
(i.e., $\widetilde{x}_2(k+1)=\widetilde{x}_2(k)$) as
\begin{equation}
 \widetilde{x}_2(k)
  = (1-m)\big[
            I - (1-m)\widetilde{A}_{22}
          \big]^{-1}
           \widetilde{A}_{21} \widetilde{x}_1(k),
 \label{eqn:updateVx_approx2}    
\end{equation}
where the matrix $I - (1-m)\widetilde{A}_{22}$ is nonsingular.
This expression is motivated by the time-scale separation in 
singular perturbation based approaches.
Substituting this into the recursion \eqref{eqn:updateVx1} 
for $\widetilde{x}_1(k)$ yields
\begin{align}
 \widetilde{x}_1(k+1)
  &= (1-m) 
     \Big\{
       \widetilde{A}_{11} 
         + (1-m)\widetilde{A}_{12}
            \big[
              I - (1-m)\widetilde{A}_{22}
            \big]^{-1}
             \widetilde{A}_{21} 
     \Big\}
       \widetilde{x}_1(k)
       + \frac{m}{n}u.
 \label{eqn:updateVx_approx1}    
\end{align}
Note that if this recursion is stable, then 
the steady states of the scheme above with 
\eqref{eqn:updateVx_approx2} and \eqref{eqn:updateVx_approx1}
become the same as those of 
\eqref{eqn:updateVx1} and \eqref{eqn:updateVx2};
they are equal to the transformed PageRank 
$\widetilde{x}^*$ in \eqref{eqn:xtilde}.

In this approximate form \eqref{eqn:updateVx_approx2} and
\eqref{eqn:updateVx_approx1}, the scheme requires the recursive 
computation of only the first state $\widetilde{x}_1(k)$, whose dimension 
equals the number $r$ of groups. It thus appears that
information should be exchanged only among groups.
However, we notice that 
the term $\widetilde{A}_{12}[I - (1-m)\widetilde{A}_{22}]^{-1}
\widetilde{A}_{21} \widetilde{x}_1(k)$ involves the
product of vectors of dimension $n-r$ and consequently 
may not be suitable for distributed computation. 

To reduce the computation and communication, we further simplify 
the scheme by relaxing the objective to that of computing the approximated
version of the state $\widetilde{x}(k)$.
Specifically, we modify the scheme \eqref{eqn:updateVx_approx2}
and \eqref{eqn:updateVx_approx1} above under the assumption that
$\delta$ is sufficiently small. 
The scheme consists of three steps and is given as follows.

\begin{algorithm}\label{alg:1}\rm
1.~Take the initial state $\widetilde{x}_1(0)\in\R^r$ as a probability vector.
At each time $k$,
compute the first state $\widetilde{x}_1(k)\in\R^r$ 
via the reduced-order recursion
\begin{align}
 \widetilde{x}_1(k+1)
  &= (1-m)\widetilde{A}_{11} \widetilde{x}_1(k)
       + \frac{m}{n}u.
\label{eqn:updateVx_red1}
\end{align}

2.~After the updates for $\widetilde{x}_1(k)$ converge,
compute the second state $\widetilde{x}_2(k)\in\R^{n-r}$ by
\begin{align}
 \widetilde{x}_2(k)
  &= (1-m)\big[
            I - (1-m)\widetilde{A}'_{22}
          \big]^{-1}
           \widetilde{A}_{21} \widetilde{x}_1(k).
\label{eqn:updateVx_red2}
\end{align}

3.~The state is transformed back in 
the original coordinate by
\begin{equation}
 x(k) = W \widetilde{x}(k) 
      = W_1 \widetilde{x}_1(k)
         + W_2 \widetilde{x}_2(k).
 \label{eqn:updateVx_red3}
\end{equation}
\end{algorithm}

In summary, we obtained Algorithm~\ref{alg:1}, which is an approximated 
version of the the scheme in \eqref{eqn:updateVx1} and \eqref{eqn:updateVx2}.
The approach in the derivation outlined above is 
(i)~to use the steady state of $\widetilde{x}_2(k)$, and then
(ii)~to assume small $\delta$ so that $\widetilde{A}_{12}$ and the
entries of $\widetilde{A}_{22}$ outside the diagonal blocks become small
(due to Lemma~\ref{lem:AE}\;(iii)).
In particular, the original scheme is triagonalized by replacing
$\widetilde{A}_{12}$ with zeros; as a result, in the first step 
of the algorithm, only the $r$-dimensional dynamics for the group 
values remains.  Moreover, in the second step, $\widetilde{A}_{22}'$
is block diagonal, so the matrix inversion in \eqref{eqn:updateVx_red2}
can be done at each group (while the first step is running).
The level of approximation is guaranteed via detailed analyses provided 
in Theorems~\ref{thm:xred} and \ref{thm:x2} in the next subsection.


The convergence of this scheme is outlined below. 
Similarly to Definition~\ref{def:pagerank} for
the original PageRank vector $x^*$, 
let $\widetilde{x}_1'\in\R^r$ be the eigenvector 
of the matrix $(1-m)\widetilde{A}_{11}+ (m/n) u\one_r^T$
corresponding to eigenvalue 1 as
\begin{equation}
  \widetilde{x}_1'
    = \big[
        (1-m)\widetilde{A}_{11}+ \frac{m}{n} u\one_r^T
      \big]
        \widetilde{x}_1',~~
     \widetilde{x}_1'\in[0,1]^r,~~\one_r^T \widetilde{x}_1' = 1.
\label{eqn:xtilde_1}
\end{equation}
This eigenvector exists and is unique because
$\widetilde{A}_{11}$ is stochastic by Lemma~\ref{lem:AE}\;(i) and
moreover, $u/n$ is a positive probability vector by definition;
hence, this matrix $(1-m)\widetilde{A}_{11}+ (m/n) u\one_r^T$
is positive stochastic and Perron's theorem \cite{HorJoh:85}
can be applied. Then, let
\begin{equation}
  \widetilde{x}'
    := \begin{bmatrix}
         \widetilde{x}_1'\\
         \widetilde{x}_2'
       \end{bmatrix},~~
    \text{where 
     $\widetilde{x}_2'
      := (1-m)\big[
                I - (1-m)\widetilde{A}'_{22}
              \big]^{-1}
               \widetilde{A}_{21} \widetilde{x}_1'$}.
   \label{eqn:xhat_dash}   
\end{equation}
The first part $\widetilde{x}_1'$ is the approximate of 
the aggregated PageRank $\widetilde{x}_1^*$; 
with some abuse of terminology, it 
will also be called the aggregated PageRank.
Finally, we transform this $\widetilde{x}'$ back to 
the original coordinate, and let 
\begin{equation}
  x' := V^{-1}\widetilde{x}'.
\label{eqn:xdash0}
\end{equation}

The update scheme in the algorithm is guaranteed to converge to 
this approximated PageRank vector $x'$.
We state this fact as a proposition, which follows 
from Lemma~\ref{lem:power1}.

\begin{proposition}\rm\label{prop:power2}
In the three-step update scheme in
\eqref{eqn:updateVx_red1}--\eqref{eqn:updateVx_red3},
for any initial vector $\widetilde{x}_1(0)$ that is a 
probability vector, 
it holds that the state $x(k)$ converges to $x'$ in \eqref{eqn:xdash0}
as $k\rightarrow\infty$.
\end{proposition}

We have a few remarks regarding this algorithm
from the viewpoint of distributed computation.
In the first step \eqref{eqn:updateVx_red1},
the $r$-dimensional state $\widetilde{x}_1(k)$ represents the group values.
This step requires exchange of states 
only among groups and not among individual pages. 
Hence, it is suitable for distributed computation. 
Once it reaches the steady state, the other two steps
should be carried out. 
The second step \eqref{eqn:updateVx_red2}
requires transmission over most links in the web 
for communicating the $(n-r)$-dimensional vector
$\widetilde{A}_{21} \widetilde{x}_1(k)$.
Nevertheless, the subsequent computation 
in this step as well as the third step \eqref{eqn:updateVx_red3}
can be done locally within each group.
This is because the matrices $I - (1-m)\widetilde{A}_{22}'$,
$W_1$, and $W_2$ are all block diagonal.

\begin{table}[t]
\begin{center}
\caption{Comparison of operation costs 
with communication among groups}
\label{table:operation}
\vspace*{1mm}
\begin{tabular}{lll}
  \hline
   Algorithm & Equation & Bound on numbers of operations\\
  \hline
  Original & \eqref{eqn:xM0} 
    & $O((2 f_{0}(A)+n)\overline{k}_1)$\\
  Aggregation based & \eqref{eqn:updateVx_red1}
    & $O((2f_{0}(\widetilde{A}_{11})+r)\overline{k}_2
            + f_{0}(A_{\text{ext}}) + n + r)$\\
                    & \eqref{eqn:updateVx_red2}
    & $O(2 f_{0}(A)+2n+r)$\\
  \hline
\end{tabular}
\end{center}
\hspace*{7cm} 
$f_0(\cdot)$: The number of nonzero entries of a matrix\\
\hspace*{7cm} 
$\overline{k}_1,\overline{k}_2$: The numbers of steps in the recursions
\end{table}

\begin{remark}\label{rem:comp}\rm
The computational advantage of the aggregation-based approach can be 
highlighted in terms of its operation cost \cite{GolVan:96,LinShiWei:09}. 
Table~\ref{table:operation} summarizes the numbers of operations
for the original scheme \eqref{eqn:xM0} and the proposed scheme 
\eqref{eqn:updateVx_red1}--\eqref{eqn:updateVx_red3} 
in Algorithm~\ref{alg:1}. 
Here, $f_{0}(A)$ denotes the number of nonzero entries
in the link matrix $A$. For a sparse matrix, its product with
a vector requires operations of order $2f_{0}(A)$.
Also, $\overline{k}_1$ and $\overline{k}_1$ are 
the numbers of steps required for the 
convergence of the recursions; for termination criteria, 
see, e.g., \cite{KamHavGol:04} for the centralized case 
and \cite{IshTem:10} for the distributed case.

For the proposed scheme, the operations that 
involve interaction among groups via communication
are shown; other steps can be done decentrally and
are of polynomial orders of $n_i$ for group $i$.
The first step \eqref{eqn:updateVx_red1}
requires the computation of $\widetilde{A}_{11}
=I+V_1A_{\text{ext}}W_1$ and the iteration. 
For the second step \eqref{eqn:updateVx_red2}, 
we counted the multiplication of $\widetilde{A}_{21}\widetilde{x}_{1}(k)$.
As we discussed earlier, here, 
the matrix $\widetilde{A}_{22}'$ is block diagonal, whose blocks
are of the size $(n_i-1)\times(n_i-1)$. The inverse of each block
can be computed by the corresponding group. 
The same holds for the third step \eqref{eqn:updateVx_red3},
where the transformation matrices $W_1$ and $W_2$ are
also block diagonal.

Later in Section~\ref{sec:example}, through a numerical example, 
we demonstrate that even with this reduced computation cost, 
the aggregation-based approach exhibits high performance in 
convergence rate and accuracy.
In particular, two distributed algorithms are compared: One with the 
original order $n$ and the other with the reduced order $r<n$. 
The results show that the errors from the true PageRank
decrease to comparable levels at similar rates. 
\End
\end{remark}

\subsection{Aggregated PageRank and its approximation error}
\label{sec:error}

In this subsection, we present two results establishing error bounds for 
the update scheme in Algorithm~\ref{alg:1}.
The results provide useful guidelines on 
how the aggregation of the web should be done.

The first theorem is based on the sparsity property 
in the graph $\Gcal$, represented by the node parameter 
in Assumption~\ref{assump:delta}.

Letting $\epsilon \in (0,1)$ be a parameter that determines the 
desired level of approximation, 
we consider the upper bound $\delta$ on node parameters.
Aggregate the web so that $\delta$ is sufficiently small that
\begin{equation}
   \delta 
       \leq \frac{m\epsilon}{4(1-m)(1+\epsilon)}.
   \label{eqn:delta2}
\end{equation}

\begin{theorem}\label{thm:xred}\rm
Under Assumption~\ref{assump:delta} with the parameter 
$\delta$ satisfying \eqref{eqn:delta2}, 
the error between the steady state $x'$ in \eqref{eqn:xdash0} 
of the update scheme Algorithm~\ref{alg:1}
and the PageRank vector $x^*$ is bounded as
\begin{equation}
   \norm{x^* - x'}_1 \leq \epsilon.
\label{eqn:thm:xred}
\end{equation}
\end{theorem}

To prove this theorem, it is useful to 
consider the scheme given by
\begin{align}
  \begin{bmatrix}
    \widetilde{x}_1(k+1)\\
    \widetilde{x}_2(k+1)
  \end{bmatrix}
    &= (1-m) 
        \widetilde{A}'
        \begin{bmatrix}
          \widetilde{x}_1(k)\\
          \widetilde{x}_2(k)
        \end{bmatrix}
       + \frac{m}{n}
          \begin{bmatrix}
            u\\
            0
          \end{bmatrix},
\label{eqn:xhat}        
\end{align}
where the matrix $\widetilde{A}'$ is defined as
\begin{equation}
  \widetilde{A}'
    := \begin{bmatrix}
         \widetilde{A}_{11} & 0\\
         \widetilde{A}_{21} & \widetilde{A}'_{22}
       \end{bmatrix}.  
\label{eqn:Atil_dash}       
\end{equation}
This is a modified version of $\widetilde{A}$ by
replacing $\widetilde{A}_{12}$ and $\widetilde{A}_{22}$
with $0$ and $\widetilde{A}_{22}'$, respectively.
Note that the matrix $(1-m)\widetilde{A}'$ is stable because 
by Lemma~\ref{lem:AE}, $\widetilde{A}_{11}$ is stochastic and
$(1-m)\widetilde{A}'_{22}$ is stable. 
It is straightforward to show that in the scheme \eqref{eqn:xhat}, 
the state converges to $\widetilde{x}'$ in \eqref{eqn:xhat_dash}.
Then, the vector $x'=V^{-1}\widetilde{x}'$ given in \eqref{eqn:xdash0}
must be such that
\begin{equation}
   x' = (1-m)A'x' + \frac{m}{n}\one_n,
     ~\text{where}~
      A' := V^{-1}\widetilde{A}'V. 
   \label{eqn:xdash}
\end{equation}

We start with a preliminary result regarding this matrix $A'$.

\begin{lemma}\label{lem:normAA}\rm
 Under Assumption~\ref{assump:delta}, it holds that
 $\norm{A-A'}_1\leq 4\delta$. 
\end{lemma}

\Proof
By the definitions of $\widetilde{A}$ and $\widetilde{A}'$,
we have $A - A'= V^{-1} (\widetilde{A} - \widetilde{A}') V$.
Using the submatrix expressions of \eqref{eqn:Atil} and
\eqref{eqn:Atil_dash} and also \eqref{eqn:A22dash}, we have
\begin{align*}
  A - A' 
    &= W_1 \widetilde{A}_{12} V_2 
        + W_2 (\widetilde{A}_{22}-\widetilde{A}'_{22}) V_2 
    = (W_1 V_1 + W_2 V_2) A_{\text{ext}} W_2 V_2
    = A_{\text{ext0}} W_2 V_2,
\end{align*}
where the last equality holds by \eqref{eqn:AE}
and $W_1 V_1 + W_2 V_2=I$.
The matrix $W_1V_1$ is block diagonal in the form of
$\text{bdiag}(1/\widetilde{n}_i 
\one_{\widetilde{n}_i}\one_{\widetilde{n}_i}^T)$
and is stochastic. Hence, $\norm{W_1V_1}_1=1$. 
Also, $W_2 V_2=I-W_1V_1$, which implies $\norm{W_2V_2}_1\leq 2$. 
Therefore, by Lemma~\ref{lem:AE}\;(ii), 
the 1-norm of $A-A'$ can be bounded as 
\begin{equation}
  \norm{A-A'}_1 
    \leq \norm{A_{\text{ext0}}}_1\;
                 \norm{W_2 V_2}_1
    \leq 4\delta.
    \tag*{\EndProof}
\end{equation}

\smallskip
{\it Proof of Theorem~\ref{thm:xred}:}~%
From \eqref{eqn:prvec} and \eqref{eqn:xdash}, it follows that
\begin{align*}
  x^* - x' 
    &= (1-m)(Ax^* - A' x')
     = (1-m)[(A-A')x^* + A' (x^*-x')].
\end{align*}
Thus, we have
$[I-(1-m)A'](x^* - x') = (1-m)(A-A')x^*$.
Here note that the matrix $I-(1-m)A'$ is nonsingular 
because $(1-m)A'$ is stable. Thus, we obtain
\begin{equation}
  x^* - x' = (1-m)[I-(1-m)A']^{-1}(A-A')x^*.
  \label{eqn:thm:xred1}
\end{equation}
By the condition \eqref{eqn:delta2} on $\delta$, 
it also holds that
\[
  (1-m)\norm{A'}_1
    \leq (1-m)[\norm{A}_1 + \norm{A'-A}_1]
    \leq (1-m)(1+4\delta)
    < 1,
\]
where the second inequality is due to Lemma~\ref{lem:normAA}.
Hence, from \eqref{eqn:thm:xred1}, we have
\begin{align*}
  \norm{x^*-x'}_1
    &\leq (1-m)
           \Big\|
             \sum_{k=0}^{\infty} (1-m)^k(A')^k 
             (A-A') x^*
           \Big\|_1\\
    &\leq (1-m) \sum_{k=0}^{\infty} [(1-m)\norm{A'}_1]^k 
           \norm{A-A'}_1\, \norm{x^*}_1\\
    &\leq \frac{4\delta(1-m)}{1-(1-m)(1+4\delta)}.
\end{align*}
Finally, by the bound \eqref{eqn:delta2} on $\delta$, 
we obtain the inequality in \eqref{eqn:thm:xred}.
\EndProof

\medskip
The theorem exhibits that aggregation is useful
in obtaining a good approximate of the PageRank 
based on an update scheme of a lower order.
For the approximate calculation to be feasible,
the critical condition is (i) of Assumption~\ref{assump:delta},
setting a limit on the ratio of external links for each group.
It is however clear that in the web,
many pages have many external links outside its own domain,
which will not satisfy this assumption.
Such a page should not be grouped with
other pages, but instead be treated as a group on its own;
these pages will then satisfy (ii) of Assumption~\ref{assump:delta}.
The proposed aggregation method is closely related to those
considered in the context of singular perturbation analyses.
The differences will be discussed in detail in Section~\ref{sec:sing}.


\medskip
We proceed to the second result, 
which also shows how
the web aggregation should be carried out from a different perspective.
In the proposed scheme, the first step \eqref{eqn:updateVx_red1}
involves only the state $\widetilde{x}_1$, which consists
of the total values of each group.
Hence, when the coordinate is transformed back 
as $x=W\widetilde{x}=W_1\widetilde{x}_1 + W_2\widetilde{x}_2$ 
in the third step \eqref{eqn:updateVx_red3},
each entry of its contribution $W_1 \widetilde{x}_1$ represents
the average value of the group to which the corresponding page 
belongs.
This means that if the grouping is done so that the values of 
the members in each group are similar, we can expect 
that the scheme computes $\widetilde{x}_1$ with small error.
The following theorem provides a quantitative 
result on this intuition. 

\begin{theorem}\label{thm:x2}\rm
If $\norm{x^* - W_1 \widetilde{x}_1^*}_1 \leq \kappa$,
then it holds that 
\begin{equation}
   \norm{x^* - W_1 \widetilde{x}'_1}_1
     \leq \frac{\kappa}{m}.
\label{eqn:thm:x2}  
\end{equation}
\end{theorem}

\Proof
By \eqref{eqn:xtilde}, we have
$x^*=W\widetilde{x}^*=W_1 \widetilde{x}_1^*+W_2 \widetilde{x}_2^*$.
Thus, 
\begin{equation}
  x^* - W_1 \widetilde{x}'_1
    = W_1 (\widetilde{x}_1^*-\widetilde{x}'_1)
       + W_2 \widetilde{x}_2^*.
\label{eqn:thm:x2:1}
\end{equation}
This means that we shall focus on $\widetilde{x}_1^*-\widetilde{x}'_1$.
Observe that by definition,
$\widetilde{x}_1^*$ is part of the equilibrium of 
the recursion in \eqref{eqn:updateVx1} and \eqref{eqn:updateVx2}. 
Also, $\widetilde{x}'_1$ is the equilibrium of 
\eqref{eqn:updateVx_red1}. Thus, it follows that
\begin{align*}
 \widetilde{x}_1^*-\widetilde{x}'_1
  = (1-m)\big[
           \widetilde{A}_{11}
             (\widetilde{x}_1^* - \widetilde{x}'_1)
            + \widetilde{A}_{12} \widetilde{x}_2^*
         \big].
\end{align*}
Hence, we have 
$\big[I-(1-m)\widetilde{A}_{11}\big]
\big(\widetilde{x}_1^*-\widetilde{x}'_1\big)
= \widetilde{A}_{12}\widetilde{x}_2^*$.
By Lemma~\ref{lem:AE}\;(i), $\widetilde{A}_{11}$ is a
stochastic matrix, and consequently 
$\rho((1-m)\widetilde{A}_{11})=1-m<1$.
This implies that $(1-m)\widetilde{A}_{11}$ is a 
stable matrix, and hence $I-(1-m)\widetilde{A}_{11}$
is nonsingular. As a result, 
it holds that
\begin{align} 
  \widetilde{x}_1^*-\widetilde{x}'_1
   &= (1-m)\big[I-(1-m)\widetilde{A}_{11}\big]^{-1}
       \widetilde{A}_{12}\widetilde{x}_2^* 
   = (1-m)\sum_{k=0}^{\infty} 
       \big[(1-m)\widetilde{A}_{11}\big]^{k}
        \widetilde{A}_{12}\widetilde{x}_2^* \notag\\
   &= (1-m)\sum_{k=0}^{\infty} 
       [(1-m)V_1 A W_1]^{k}
        V_1 A W_2 \widetilde{x}_2^*,
\label{eqn:thm:x2:3}
\end{align}
where in the last equality we used \eqref{eqn:Atil}.
Substitution of \eqref{eqn:thm:x2:3} into \eqref{eqn:thm:x2:1} 
results in
\begin{align*} 
  W_1 \big(
        \widetilde{x}_1^*-\widetilde{x}'_1
      \big)
        + W_2 \widetilde{x}_2^*
   &= \sum_{k=1}^{\infty} 
       \big\{
          [(1-m)(W_1 V_1 A)]^{k}
           + I
       \big\}
         W_2 \widetilde{x}_2^*.
\end{align*}
Now it follows that
\begin{align*}
 \bigl\|
    W_1 \big(
          \widetilde{x}_1^*-\widetilde{x}'_1
        \big)
     + W_2 \widetilde{x}_2^*
 \bigr\|_1
 &\leq \biggl\{
       \biggl\|
         \sum_{k=1}^{\infty} 
            [(1-m)(W_1 V_1 A)]^{k}
       \biggr\|_1 + 1
       \biggr\}
        \bigl\|
          W_2 \widetilde{x}_2^*
        \bigr\|_1 \notag\\
 &= \biggl[
       \sum_{k=1}^{\infty} 
          \bigl\|
           [(1-m)(W_1 V_1 A)]^{k}
          \bigr\|_1 + 1
       \biggr]
        \bigl\|
          W_2 \widetilde{x}_2^*
        \bigr\|_1\\
 &\leq \biggl[
       \sum_{k=1}^{\infty}  (1-m)^k
           + 1
    \biggr]\kappa
   = \frac{\kappa}{m},
\end{align*}
where the first equality holds since 
$W_1 V_1 A$ is a stochastic matrix
and the second inequality is due to the condition
$\norm{W_2 \widetilde{x}_2^*}_1
=\norm{x^*-W_1 \widetilde{x}_1^*}_1\leq \kappa$.
Therefore, we arrive at the bound in \eqref{eqn:thm:x2}.
\EndProof

\medskip
The condition $\norm{x^*-W_1 \widetilde{x}_1^*}_1\leq \kappa$ 
in the theorem
may in general be difficult to check
because it requires global information about PageRank.
However, it is possible to convert it to a local condition.
In fact, a sufficient condition is that,
for each page $i\in\Vcal$, 
the relative error between the value $x^*_i$ and the average
$(W_1\widetilde{x}_1^*)_i$ of its group satisfies 
$\abs{x_i^* - (W_1\widetilde{x}_1^*)_i }\leq \kappa x_i^*$.
Obviously, this relative error is zero for 
any group with only one member. 
Thus, for more accurate computations, we may envision
to run an algorithm estimating the local value 
$\abs{x_i^*- (W_1 \widetilde{x}_1(k))_i}$
in real time; if the estimate exceeds a given
threshold, then the group should be split into smaller groups,
each of which having smaller relative errors.
On the other hand, the theorem is stated in terms of the 1-norm 
of the approximation errors.
As we see in the proof, for this particular norm, 
a fairly tight bound is obtained;
the reason is that 
for column stochastic matrices, the 1-norm is always 1.


\begin{figure}[t]
  \centering
  \vspace*{4mm}
  \fig{7.2cm}{!}{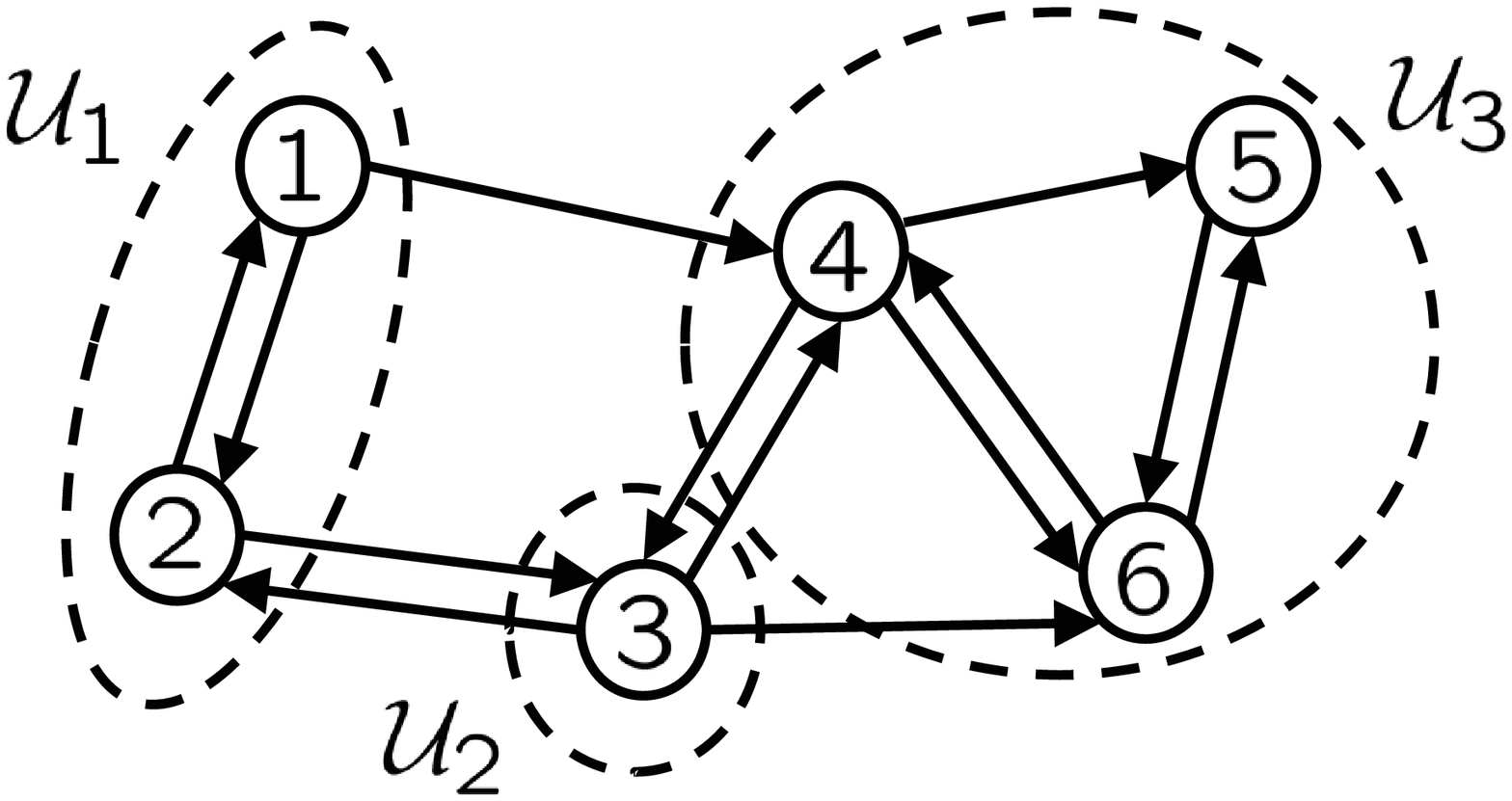}
  \caption{The example web, where the dashed lines indicate the
           grouping for aggregation}
  \label{fig:graph}
\end{figure}

\subsection{Example}
\label{ex:orig}

We present a simple example to illustrate 
the idea of PageRank via web aggregation.

Consider the web consisting of six pages shown 
in Fig.~\ref{fig:graph}. As a graph, this web is strongly connected. 
The original link matrix $A$ in \eqref{eqn:xA:pr}
is given by
%
\begin{equation*}
 A = \begin{bmatrix}
       0 & 1/2 & 0 & 0 & 0 & 0\\
       1/2 & 0 & 1/3  & 0 & 0 & 0\\
       0 & 1/2 & 0 & 1/3 & 0 & 0\\
       1/2 & 0 & 1/3 & 0 & 0 & 1/2\\
       0 & 0 & 0 & 1/3 & 0 & 1/2\\
       0 & 0 & 1/3 & 1/3 & 1 & 0
     \end{bmatrix}.
\label{eqn:ex:orig}
\end{equation*}
The PageRank vector $x^*$ in \eqref{eqn:prvec}
can be found as
\begin{equation}
  x^* = \begin{bmatrix}
           0.0614 & 0.0857 & 0.122 & 0.214 & 0.214 & 0.302
        \end{bmatrix}^T.
\label{eqn:ex:x_ast}        
\end{equation}
Pages~4 and 6 have the largest number of incoming links,
resulting in large PageRank values.
Page~6 is
more advantageous because the pages contributing to its value
via links, i.e., pages~3, 4, and 5, 
have larger values than those having links to page~4. 
In particular, page~1 has the smallest number
of incoming links and obviously the lowest ranking in this web. 

\begin{figure}
  \vspace*{2mm}
  \centering
  \fig{5.2cm}{!}{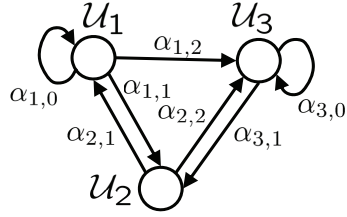}  
  \vspace*{-3mm}
  \caption{Aggregated graph 
           with the update probabilities $\alpha_{i,\ell}$ 
           for the links (see also Example~\ref{ex:agg3})}
  \label{fig:agg}  
\end{figure}

We now aggregate the web and partition the nodes into three
groups (i.e., $r=3$) as 
$\Ucal_1=\{1,2\}$, $\Ucal_2=\{3\}$, and $\Ucal_3=\{4,5,6\}$.
These are indicated by the dashed lines in Fig.~\ref{fig:graph}, and
the aggregated graph is illustrated in Fig.~\ref{fig:agg}. 
The nodes $\Ucal_1$ and $\Ucal_3$ contain self-loops while
$\Ucal_3$ does not since it is a single group.
Here, the node parameters are
$\delta_1 = \delta_2 = 1/2$, $\delta_3 = 1$, 
$\delta_4 = 1/3$, and $\delta_5 = \delta_6 = 0$.
Thus, all the pages satisfy Assumption~\ref{assump:delta}
by taking $\delta=0.5$.
This grouping is also reasonable from the viewpoint of
Theorem~\ref{thm:x2}
because in the true PageRank vector $x^*$, the values of the pages 
in the groups $\Ucal_1$ and $\Ucal_3$ are relatively close.
In this case,
the transformation matrix $V$ in \eqref{eqn:V} is
%
\[
 V = \left[
       \begin{array}{c}
         V_1\\       
         \hline
         V_2
       \end{array}
     \right]
   = \left[
     \begin{array}{ccccccc}
       \cdashline{1-2}
       \multicolumn{1}{:c}{1} & \multicolumn{1}{c:}{1} & 0 & 0 & 0 & 0\\
       \cdashline{1-3}
       0 & 0 & \multicolumn{1}{:c:}{1} & 0 & 0 & 0\\
       \cdashline{3-6}
       0 & 0 & 0 & \multicolumn{1}{:c}{1} & 1 & \multicolumn{1}{c:}{1}\\
       \cdashline{4-6}\\[-4mm]
       \hline\\[-4mm]
       \cdashline{1-2}
       \multicolumn{1}{:c}{1/2} & \multicolumn{1}{c:}{-1/2} & 0   & 0 & 0 & 0\\
       \cdashline{1-2}\cdashline{4-6}
       0 & 0 & 0 &  \multicolumn{1}{:c}{2/3} & -1/3 & \multicolumn{1}{c:}{-1/3}\\
       0 & 0 & 0 & \multicolumn{1}{:c}{-1/3} &  2/3 & \multicolumn{1}{c:}{-1/3}\\
       \cdashline{4-6}
    \end{array}\right],
\]
%
where the diagonal blocks are indicated by dashed-line boxes.
In $V_2$, the third column corresponding to the single group $\Ucal_2$
is zero. 
Then, the PageRank after the coordinate transformation, 
$\widetilde{x}^*=V x^*$, can be found as
\[
 \widetilde{x}^* 
  = \left[
       \begin{array}{c|c}
         (\widetilde{x}_1^*)^T & (\widetilde{x}_2^*)^T  
       \end{array}
    \right]^T
  = \left[
       \begin{array}{ccc|ccc}
         0.147 & 0.122 & 0.731 & -0.0121 &  -0.0294 & -0.0294
       \end{array}
    \right]^T.    
\]
Notice that the first state $\widetilde{x}_1^*$ is a probability vector.

For this grouping, by \eqref{eqn:Adecomp}, 
the matrix $A$ can be decomposed as
$A = I + A_{\text{int}} + A_{\text{ext}}$, where
the internal matrix $A_{\text{int}}$ and 
the external matrix $A_{\text{ext}}$ are respectively
%
\begin{align*}
 A_{\text{int}} 
  = \left[
     \begin{array}{cccccc}
      \cdashline{1-2}
      \multicolumn{1}{:c}{-1/2} & \multicolumn{1}{c:}{1/2}  &  0  & 0 &  0  &  0\\
      \multicolumn{1}{:c}{1/2}  & \multicolumn{1}{c:}{-1/2} &  0  & 0 &  0  &  0\\
      \cdashline{1-3}
         0 &  0   &  \multicolumn{1}{:c:}{0}  & 0 &  0  &  0\\
      \cdashline{3-6}
         0 &  0   &  0  & \multicolumn{1}{:c}{-2/3} & 0 & \multicolumn{1}{c:}{1/2}\\
         0 &  0   &  0  & \multicolumn{1}{:c}{1/3} & -1 & \multicolumn{1}{c:}{1/2}\\
         0 &  0   &  0  & \multicolumn{1}{:c}{1/3} &  1 & \multicolumn{1}{c:}{-1} \\
      \cdashline{4-6}
     \end{array}
    \right],~~
 A_{\text{ext}}
  = \left[
     \begin{array}{cccccc}
      \cdashline{1-2}
      \multicolumn{1}{:c}{-1/2} &  \multicolumn{1}{c:}{0}  &  0   &  0   &  0  &  0\\
      \multicolumn{1}{:c}{0}   & \multicolumn{1}{c:}{-1/2} &  1/3 &  0   &  0  &  0\\
      \cdashline{1-3}
       0   & 1/2 & \multicolumn{1}{:c:}{-1}   &  1/3 &  0  &  0\\   
      \cdashline{3-6}
       1/2 &  0   &  1/3 & \multicolumn{1}{:c}{-1/3} &  0  & \multicolumn{1}{c:}{0}\\   
         0 &  0   &  0   & \multicolumn{1}{:c}{0}    &  0  & \multicolumn{1}{c:}{0}\\
         0 &  0   &  1/3 & \multicolumn{1}{:c}{0}    &  0  & \multicolumn{1}{c:}{0}\\
      \cdashline{4-6}
     \end{array}
    \right].
\end{align*}
%
The internal matrix $A_{\text{int}}$ consists of the block-diagonal 
elements of $A$ while the external matrix $A_{\text{ext}}$
contains the rest.

The link matrix $\widetilde{A}$ in \eqref{eqn:Atil0}
in the corresponding coordinate becomes
%
\begin{align*}
  \widetilde{A}
    &= \left[
         \begin{array}{c|c}
          \widetilde{A}_{11} & \widetilde{A}_{12}\\
          \hline
          \widetilde{A}_{21} & \widetilde{A}_{22}
         \end{array}
       \right]
    = \left[
         \begin{array}{ccc|ccc}
           0.5  & 0.333 & 0     &  0   &  0     & 0\\
           0.25 & 0     & 0.111 & -0.5 &  0.333 & 0\\
           0.25 & 0.667 & 0.889 &  0.5 & -0.333 & 0\\
           \hline
           0   & -0.167 &      0 &  -0.5  &        0  &       0\\
           0.167 &  0.111 & -0.130 &   0.333 &   -0.389 &   -0.5\\   
          -0.0833 &  -0.222 &  -0.0185 & -0.167 &   -0.0556 &   -0.5
	     \end{array}
	  \right].
\end{align*}
%
The proposed update scheme 
in Algorithm~\ref{alg:1}
employs the matrices $\widetilde{A}_{11}$ above and
%
\begin{align*}
  \begin{bmatrix}
    I - (1-m)\widetilde{A}'_{22}
  \end{bmatrix}^{-1}
    \widetilde{A}_{21}
     = \begin{bmatrix}
         0 & -0.167 & 0\\
         0.174 & 0.161 & -0.113\\
         -0.0758 & -0.172 & -0.00177
       \end{bmatrix},
\end{align*}
%
where $\widetilde{A}'_{22}$ is obtained through
\eqref{eqn:A22dash} using $A_{\text{int}}$ 
%
\begin{align*}
  \widetilde{A}'_{22}   
    = I + V_2 A_{\text{int}} W_2
    = \left[\begin{array}{ccc}
        \cdashline{1-1}
        \multicolumn{1}{:c:}{0} &   0     &   0\\
        \cdashline{1-3}
        0 &  \multicolumn{1}{:c}{-0.167} &  \multicolumn{1}{c:}{-0.5}\\
        0 &  \multicolumn{1}{:c}{-0.167} &  \multicolumn{1}{c:}{-0.5}\\
        \cdashline{2-3}
      \end{array}\right].
\end{align*}
%
Notice that the matrix $\widetilde{A}_{11}$ is stochastic.
Also, $\widetilde{A}_{22}'$ has a block-diagonal structure
(different from $\widetilde{A}_{22}$) and is a stable matrix.
For this scheme, the steady state in the original coordinate is 
\[
  x' = W \widetilde{x}'
     = \begin{bmatrix}
         0.0566 & 0.0920 & 0.125 & 0.212 & 0.213 & 0.302
       \end{bmatrix}^T.
\] 
Comparing this with the true value $x^*$ in \eqref{eqn:ex:x_ast}, 
the error is indeed small as
$\norm{x'-x^*}_1 
    = 0.0188$.

\section{Discussion on aggregation-based methods}
\label{sec:sing}

In this section, we provide some discussion on 
our results from the viewpoint of aggregation.
As mentioned in the Introduction, the approach of this 
paper has been motivated by the singular 
perturbation results of \cite{AldKha:91,ChoKok:85} 
for Markov chains and \cite{BiyArc:08,PhiKok:81} for consensus-like 
problems with sparse network structures.
We have several remarks in relation to these works. 

In the Markov chain literature (e.g.\ \cite{MeyTwe:09}), 
the problem of finding the stationary probability distribution 
based on aggregation has been long studied; see, for example, \cite{Courtois:75}. 
The paper \cite{Meyer:89} formalizes a general method
for finding the exact distribution,
and its application to the PageRank computation is 
discussed in \cite{LanMey:06}.
The approach of \cite{PhiKok:81} can be seen as an interpretation 
of this method from the viewpoint of singular perturbation.

The papers \cite{AldKha:91,PhiKok:81} consider the 
special case when
the chain has the so-called \textit{nearly completely 
decomposable} structure. In the context of our paper, 
this means that the external link matrix $A_{\text{ext}}$
in \eqref{eqn:Adecomp} can be bounded as $\norm{A_{\text{ext}}}_1\leq \epsilon'$
with a small $\epsilon'>0$
so that the interaction among different groups is weak. 
If $\epsilon'$ is sufficiently small, 
the recursion \eqref{eqn:updateVx1} and \eqref{eqn:updateVx2} can 
be transformed to the singular perturbation form, to which
standard results (e.g., \cite{Khalil:96}) can be applied. 
Note that in Lemma~\ref{lem:AE}\;(iii), the bound on the external 
link matrix is for $A_{\text{ext0}}$ and not $A_{\text{ext}}$.

Similarly, the works of \cite{BiyArc:08,ChoKok:85} deal with problems
for multi-agent systems on consensus.
The specific setup involves
undirected graphs and hence a link matrix which 
is symmetric and stochastic (i.e., not only column 
stochastic as in the PageRank problem).
In these papers, 
simple transformation matrices similar to $V$ in \eqref{eqn:V} have
been used.
In their results, assumptions are made on the node parameters for
\textit{all} pages and also on the average value of the 
node parameters for each group; these can be (roughly) stated as 
$\delta_i\leq \delta$ for all $i$ and 
$\sum_{i\in\Ucal_j}\delta_i\leq \epsilon'$ for all $j$ 
with $\delta,\epsilon'>0$.
The consequence is that $\widetilde{A}_{11}$ and 
$\widetilde{A}_{21}$ can be bounded by constant multiples 
of $\epsilon'$ and hence the problem becomes similar to the
Markov chain case mentioned above. 

It however is important to note that the web may not have such
strong sparsity properties as those assumed in the abovementioned works.
For instance, for a page belonging to a small group,
one external link can result in a large node parameter.
By contrast, the assumption imposed in our approach 
is the condition 
$\delta_i\leq \delta$ only in the case page $i$ belongs to a 
group consisting of multiple members; this condition can be
checked easily in the grouping procedure outlined in Remark~\ref{rem:group}.
Thus, the results are applicable to a graph with any structure
after appropriately grouping the pages. 
One feature here is the tradeoff 
between the number $r$ of groups and the node parameter $\delta$
as discussed in Remark~\ref{rem:group};
from Theorem~\ref{thm:xred}, we observe that
more accurate computation requires a larger number of groups,
and thus a smaller $\delta$.

Furthermore, it is emphasized that in the aggregated
recursion \eqref{eqn:updateVx_red1}, the link matrix $\widetilde{A}_{11}$
is a stochastic matrix. This is critical in 
the distributed algorithm in the next section. 
In contrast, in the singular perturbation
form, the corresponding matrix is not necessarily stochastic
\cite{AldKha:91}.

\section{Distributed randomized algorithm for aggregated PageRank}
\label{sec:distributed}

In this section, we construct a distributed randomized scheme 
for finding the aggregated PageRank in the first step 
of Algorithm~\ref{alg:1}.

To simplify the notation, 
we rewrite the aggregated PageRank in \eqref{eqn:xhat_dash}
as $\xi':=\widetilde{x}_1'$
and moreover the recursion in the first step 
\eqref{eqn:updateVx_red1} as
\begin{equation}
 \xi(k+1)
   = (1-m) \Phi \xi(k) + \frac{m}{n} u,
\label{eqn:xi}
\end{equation}
where the link matrix is denoted by $\Phi=(\phi_{ij}):=\widetilde{A}_{11}$ and 
the state by $\xi(k):=\widetilde{x}_1(k)$. 
%
%

The objective is to compute the aggregated PageRank 
$\xi'$ via the distributed update scheme of \eqref{eqn:xi}
in the form given by
\begin{equation}
 \xi(k+1)
   = (1-\hat{m}) \Phi_{\eta(k)} \xi(k) 
         + \frac{\hat{m}}{n} u,
\label{eqn:xk2}
\end{equation}
where $\xi(k)\in\R^{r}$ is the state whose
initial condition $\xi(0)$ is a probability vector,
and $\hat{m}\in(0,1)$;
the process $\eta(k):=[\eta_1(k)\;\cdots\;\eta_{r}(k)]$ defined in 
\eqref{eqn:eta} determines the communication pattern at time $k$.
In this scheme, each group $i$ also computes the time
average of its own state $\xi_i$. Let $\psi(k)$ be
the average of $\xi(0),\ldots,\xi(k)$ as
\begin{equation}
 \psi(k) 
    = \frac{1}{k+1} \sum_{\ell=0}^{k} \xi(\ell)
    = \frac{1}{k+1} \bigl(
                       k \psi(k-1)+ \xi(k)
                    \bigr).
\label{eqn:xk2_psi}
\end{equation}

Let $\alpha\in(0,1]$, which is called the \textit{base probability}.
Recall that the update probability $\alpha_{i,\ell}$ 
in \eqref{eqn:alpha_agg} and \eqref{eqn:alpha_il}
determines the probability that group $i$ transmits
to its neighbors belonging to $\widetilde{\Vcal}_{i,\ell}$ for $\ell\neq 0$.
Assume that they are chosen so that
\begin{equation}
  \alpha_{i,\ell} 
    \in \biggl[
           \alpha \sum_{j\in\widetilde{\Vcal}_{i,\ell}} \phi_{ji}, 1
        \biggr],~~\ell=1,2,\ldots,g_i,~~
   \sum_{\ell=0}^{g_i} \alpha_{i,\ell}  
      = 1.
\label{eqn:alpha_il2}
\end{equation}
It is noted that by Lemma~\ref{lem:AE}\;(i), 
the link matrix $\Phi=\widetilde{A}_{11}$ is stochastic,
and thus $\sum_{j=1}^r \phi_{ji}=1$.
Hence, in this condition,
the lower bound 
$\alpha \sum_{j\in\widetilde{\Vcal}_{i,\ell}} \phi_{ji}$
on $\alpha_{i,\ell}$ is at most $\alpha$.

In \eqref{eqn:xk2}, 
the distributed link matrices $\Phi_{q_1,\ldots,q_{r}}$ 
for $q_{i}\in\{0,1,\ldots,g_{i}\}$, $i\in\widetilde{\Vcal}$, 
are given by 
\begin{equation}
  (\Phi_{q_1,\ldots,q_{r}})_{pi}
   := \begin{cases}
       \frac{\alpha}{\alpha_{i,\ell}} \phi_{pi} 
           & \text{if $q_i=\ell\neq 0$, $p\in\widetilde{\Vcal}_{i,\ell}$},\\
        1 - \frac{\alpha}{\alpha_{i,\ell}}
                 \sum_{j\in\widetilde{\Vcal}_{i,\ell}} \phi_{ji}                                 
           & \text{if $q_i=\ell\neq 0$, $p=i$},\\
       1   & \text{if $q_i=0$, $p=i$},\\
       0   & \text{otherwise}
     \end{cases}
\label{eqn:A_agg1}
\end{equation}
for $p,i\in\widetilde{\Vcal}$.
Notice that these link matrices are in accordance with the 
communication pattern specified by $\eta(k)$, i.e.,
$(\Phi_{\eta(k)})_{pi}>0$ if and only if 
group $i$ sends its value to group $p$
at time $k$.

Then, we can establish some desired properties of the 
link matrices for the update scheme \eqref{eqn:xk2} to converge.
These facts are stated in the proposition below.

\begin{proposition}\label{prop:Aq}\rm
 For the distributed link matrices $\Phi_{q}$ in 
 \eqref{eqn:A_agg1}, the following two properties are satisfied:
 \begin{enumerate}
   \item[(i)] For each $q$, the matrix $\Phi_{q}$ is stochastic.
   \item[(ii)] 
      The average matrix $\overline{\Phi}:=E[\Phi_{\eta(k)}]$ 
      can be written as
      $\overline{\Phi}=\alpha \Phi + (1-\alpha) I$.
 \end{enumerate}
\end{proposition}

\Proof
(i) Let $\phi_i\in\R^{r}$ be the $i$th column of $\Phi$ 
and further let
$\phi_i^{(\ell)}\in\R^{r}$ be a vector containing only those elements
corresponding to the nodes in $\widetilde{\Vcal}_{i,\ell}$, i.e.,
$\bigl(
     \phi_i^{(\ell)}
  \bigr)_p
   := \phi_{pi}$ if $p\in\widetilde{\Vcal}_{i,\ell}$ and 
$0$ otherwise
for $\ell=1,2,\ldots,g_j$ and $p,i\in\widetilde{\Vcal}$.
Notice that 
\begin{equation}
  \phi_i
     = \sum_{\ell=1}^{g_i} \phi_i^{(\ell)}.
 \label{eqn:prop:Aq:ai}
\end{equation}
Now, the $i$th column of $\Phi_{q}$ depends only on $q_i$;
so denoting this column by $\widetilde{\phi}_i^{(\ell)}$ with $\ell=q_i$, 
we have by \eqref{eqn:A_agg1}
\[
  \widetilde{\phi}_i^{(\ell)}
   = \begin{cases}
      \frac{\alpha}{\alpha_{i,\ell}} \phi_i^{(\ell)}
        + \biggl(
            1 - \frac{\alpha}{\alpha_{i,\ell}}
                   \sum_{j\in\widetilde{\Vcal}_{i,\ell}} \phi_{ji}
          \biggr) e_i &
             \text{if $\ell\in\{1,\ldots,g_i\}$},\\
      e_i  &  \text{if $\ell=0$},
     \end{cases}
\]
where $e_j\in\R^{r}$ is the unit vector 
whose $j$th element is 1 and the rest are 0.
It is now clear that $\widetilde{\phi}_i^{(\ell)}\geq 0$ because
by definition $\phi_i^{(\ell)}\geq 0$ and 
$1 - \alpha/\alpha_{i,\ell}%
\sum_{j\in\widetilde{\Vcal}_{i,\ell}} \phi_{ji}\geq 0$
by the choice of $\alpha_{i,\ell}$ in \eqref{eqn:alpha_il2}.
Moreover, if $\ell\neq 0$, it follows that 
\begin{align*}
  \norm{\widetilde{\phi}_i^{(\ell)}}_1
   = \frac{\alpha}{\alpha_{i,\ell}} \norm{\phi_i^{(\ell)}}_1
        + \biggl(
             1 - \frac{\alpha}{\alpha_{i,\ell}}
               \sum_{j\in\widetilde{\Vcal}_{i,\ell}} \phi_{ji}
          \biggr) 
   = 1,
\end{align*}
where we used the fact 
$\norm{\phi_i^{(\ell)}}_1=%
\sum_{j\in\widetilde{\Vcal}_{i,\ell}} \phi_{ji}$.
It thus follows that each column of $\Phi_{q}$ 
is nonnegative and the sum of the elements equals one; 
this implies that this matrix $\Phi_{q}$ is stochastic.

(ii)~Let $\overline{\phi}_i$ be the $i$th column of the
average matrix $\overline{\Phi}$.
By the distribution of $\eta(k)$,
it holds that
\begin{align*}
  \overline{\phi}_i
   &= E\bigl[
          \widetilde{\phi}_i^{(\eta_i(k))}
       \bigr]
   = \sum_{\ell=0}^{g_i} \alpha_{i,\ell}\widetilde{\phi}_i^{(\ell)} 
   = \alpha
        \sum_{\ell=1}^{g_i} 
           \biggl(
              \phi_i^{(\ell)}
                 - \sum_{j\in\widetilde{\Vcal}_{i,\ell}} \phi_{ji} e_i
           \biggr)
         + e_i.
\end{align*}
By \eqref{eqn:prop:Aq:ai} and stochasticity of $\Phi$,
we have $\overline{\phi}_i = \alpha \phi_i + (1-\alpha) e_i$.
This holds for $i=1,\ldots,r$, and consequently
we obtain $\overline{\Phi}=\alpha \Phi + (1-\alpha) I$.
\EndProof

\medskip
We next show that the aggregated PageRank vector $\xi'$ can be expressed 
in terms of the distributed link matrices. 
Recall that by definition, $\xi'=\widetilde{x}_1'$. 
From \eqref{eqn:xtilde_1}, we have that
$\xi'$ is the unique eigenvector of the 
the link matrix given by
$\Gamma:=(1-m)\Phi+(m/n)u\one_r$ with the property $\one_r^T\xi'=1$. 
Due to the proposition above, this characterization can be extended
using the distributed link matrices $\Phi_{\eta(k)}$. 
Define the modified link matrices by $\Gamma_{\eta(k)}
     := (1-\hat{m})\Phi_{\eta(k)} + (\hat{m}/n) u \one_r^T$,
and their average by
\[
  \overline{\Gamma} 
    := E[\Gamma_{\eta(k)}] 
     = (1-\hat{m})\overline{\Phi} + \frac{\hat{m}}{n}u\one_r^T.
\]
Take the parameter $\hat{m}$ as 
\begin{equation}
  \hat{m}
    := \frac{m \alpha}{1-(1-\alpha)m}.
\label{eqn:mhat}
\end{equation}

The following lemma is the aggregated version of 
Lemma~3.3 in \cite{IshTem:10}.

\begin{lemma}\label{lem:Mbar}\rm
For the parameter $\hat{m}$ given 
in \eqref{eqn:mhat}, we have the following:
\begin{enumerate}
\item[(i)] $\hat{m}\in(0,1)$ and $\hat{m}<m$.
\item[(ii)] 
$\overline{\Gamma} 
 = \frac{\hat{m}}{m} \Gamma 
     + \left(1 - \frac{\hat{m}}{m}\right)I$.
\item[(iii)]
The aggregated PageRank vector $\xi'=\widetilde{x}_1'$ in \eqref{eqn:xtilde_1} is 
the unique eigenvector of the average matrix $\overline{\Gamma}$ 
corresponding to the eigenvalue 1. 
\end{enumerate}
\end{lemma}

\begin{remark}\label{rem:mhat}\rm
It is interesting that the choice of $\hat{m}$ in \eqref{eqn:mhat}
is different from $m$, but is critical in establishing the lemma.
In particular, noting that the average matrix $\overline{\Gamma}$
is stochastic, we can guarantee that the average state 
$\overline{\xi}(k):=E[\xi(k)]$ of the update scheme in \eqref{eqn:xk2}
converges to the desired vector $\xi'$. This is because it follows the
recursion 
\begin{equation}
  \overline{\xi}(k+1)
    = \overline{\Gamma}\,\overline{\xi}(k),
    \label{eqn:xi_bar}
\end{equation}    
where $\overline{\xi}(0)$ is a probability vector.
We however emphasize that the state $\xi(k)$ itself does not converge
to the aggregated PageRank $\xi'$. To resolve this issue, it turns out 
to be essential to introduce the time average $\psi(k)$ as we see next. 
\End
\end{remark}

We are in the position to derive a convergence
result for the distributed scheme \eqref{eqn:xk2} 
under the probability allocation in \eqref{eqn:alpha_agg}
for the linked nodes.

\begin{theorem}\label{thm:erg2}\rm
Consider the distributed update scheme in \eqref{eqn:xk2}
and \eqref{eqn:xk2_psi}.
For any update probabilities 
$\alpha_{i,\ell}\in(0,1]$, $i\in\widetilde{\Vcal}$,
$\ell\in\{0,1,\ldots,g_i\}$,
satisfying the conditions in 
\eqref{eqn:alpha_il2},
the aggregated PageRank $\xi'$ can be obtained from 
the time average $\psi(k)$ of the states $\xi(k)$ in the mean-square sense
as $E\bigl[
   \bigl\|
      \psi(k) - \xi'
   \bigr\|^2
  \bigr] \rightarrow 0$, $k\rightarrow\infty$.
\end{theorem}

The proof of this theorem follows from noticing that, 
in Theorem~3.4 of \cite{IshTem:10}, 
to establish the mean-square convergence, the properties 
in Proposition~\ref{prop:Aq} are sufficient. 
In other words, for convergence, only stochasticity of the link 
matrices and the average behavior of the update scheme are relevant.
The type of convergence guaranteed by the theorem is
known as ergodicity for stochastic processes \cite{PapPil:02}.
While the general results of \cite{Cogburn:86} can be applied for the proof,
we have developed in \cite{IshTem:10} a more specific one, which
have been useful in extending the algorithm to incorporate 
a stopping criterion there. It was also shown there that 
the convergence rate is of order $1/k$ due to the time averaging.

The distributed update scheme presented above
has the following features:
(i)~The computation performed at each group $i$ includes
the updates in the state $\xi_i$ in \eqref{eqn:xk2}
and the time average $\psi_i$ in \eqref{eqn:xk2_psi}.
(ii)~The communication among the groups is local in the sense
that each group communicates only over direct outgoing links,
as seen from the link matrices in \eqref{eqn:A_agg1}.
(iii)~The amount of communication is determined by 
the process $\eta$, which specifies the pattern in the interaction 
between the pages. 
(iv)~At any group, the update probabilities $\alpha_{i,\ell}$
can be allocated to linked groups locally 
without information exchange among groups;
one global parameter is $\alpha$, which 
is critical for the convergence of the proposed algorithm%
\footnote{%
Practical issues related to implementation of the scheme
are outside the scope of this paper. 
Clearly, for the PageRank values reported by page owners to 
be trusted, some regulations must be enforced.
Also, reliability of the rankings can be affected by
links purposefully added to increase PageRank of certain pages; 
some works have reported
methods to detect such \textit{web spamming}
(e.g., \cite{AndBorCha:07,LanMey:06}).
Further discussions on this point are given in the footnote 
of the Introduction.}.

\begin{remark}\label{rem:links}\rm
Compared to the original scheme in \cite{IshTem:10},
a significant advantage of the one above is that the nodes need 
to communicate with neighbors only over outgoing links.
The identity of such links is necessarily contained in their local data.
This is observed in the link matrices $\Phi_q$ in \eqref{eqn:A_agg1}
where only the columns of $\Phi$ (and not the rows)
corresponding to the indices $q_i$ taking values 1 are used.
In contrast, in \cite{IshTem:10}, the protocol is that the nodes
send data over incoming links as well. 
Consequently, we also stress that the implementation of this algorithm
is simple. A closer look at the matrices $\Phi_q$ suggests us
that group $j$ should just send its current value 
with some weight to the linked groups, where
their values are updated by simply adding up the values 
received at the time. Hence, no memory is needed for 
storing values of other groups. 
This characteristic is realized by the assumption
on stochasticity of the distributed link matrices.
In contrast, such memory is necessary in algorithms
using asynchronous iteration 
\cite{BerTsi:89,deJBra:07,KolGalSzy:06}. In fact, the
most recently received values of all groups having links
to the group need to be stored. Hence, the memory size
is determined by the number of incoming links
and may be large for popular groups. 
\End
\end{remark}

The convergence rate of this scheme can be discussed from
the viewpoint of its average dynamics. As mentioned in
Remark~\ref{rem:mhat}, the average state $\overline{\xi}(k)$
converges to the aggregated PageRank $\xi'$.
Because of the recursion \eqref{eqn:xi_bar},
the asymptotic rate of convergence is exponential and
is dominated by the the second largest eigenvalue $\lambda_2(\overline{\Gamma})$ 
of $\overline{\Gamma}$ in magnitude. 
By Lemma~\ref{lem:Mbar}\;(ii), this eigenvalue can be bounded as
\begin{align}
  \abs{\lambda_2(\overline{\Gamma})}
    &= \frac{\hat{m}}{m}\abs{\lambda_2(\Gamma)} 
         + 1 - \frac{\hat{m}}{m}
    \leq \frac{1 - m}{1 - (1 - \alpha)m},
\label{eqn:lambda2Gamma}
\end{align}
where the inequality holds because 
$\abs{\lambda_2(\Gamma)} \leq 1-m$; see \cite{LanMey:06}.
Therefore, the convergence rate depends on 
the base probability $\alpha$ in communication,
i.e., more communication implies faster convergence.
On the other hand, it is interesting to observe 
that the bound \eqref{eqn:lambda2Gamma}
above is independent of the choices of the individual update probabilities
$\alpha_{i,\ell}$ as well as the number $r$ of groups. 

For the update probabilities $\alpha_{i,\ell}$ to satisfy the
conditions \eqref{eqn:alpha_il2}, 
one possible choice is the following:
\begin{equation}
  \alpha_{i,\ell}
   = \begin{cases}
       1-\alpha & \text{if $\ell=0$},\\
       \alpha 
         \frac{\sum_{j\in\widetilde{\Vcal}_{i,\ell}} \phi_{ji}}{%
               \sum_{j\in\widetilde{\Vcal}_{i}} \phi_{ji}}
                & \text{if $\ell=1,\ldots,g_i$},
     \end{cases}~~\text{for $i\in\widetilde{\Vcal}_{i}$}.
  \label{eqn:alpha_il3}
\end{equation}
In this case, the probability for group $i$ to transmit information
to some neighbor is in total equal to the base probability:
$\sum_{\ell=1}^{g_i}\alpha_{i,\ell}=\alpha$. 
Further, the probability for communicating
with group in $\widetilde{\Vcal}_{i,\ell}$ is 
proportional to the weights $\sum_{j\in\widetilde{\Vcal}_{i,\ell}} \phi_{ji}$
of the corresponding entries 
in the $i$th column of the link matrix $\Phi$.
Hence, the frequency of communication among groups with more
links is higher. 

\begin{example}\label{ex:agg3}\rm
We study the distributed algorithm based on the
aggregated web in Fig.~\ref{fig:agg} from Section~\ref{ex:orig}.
Let the communication be such that 
each group communicates with the neighbors separately and 
the update probabilities $\alpha_{i,\ell}$ are chosen as in 
\eqref{eqn:alpha_il3} above. 
Thus, for example, group~1 has two neighbor sets given by
$\widetilde{\Vcal}_{1,1}=\{2\}$ and $\widetilde{\Vcal}_{1,2}=\{3\}$.
Their update probabilities will become $\alpha_{1,1}=\alpha_{1,2}=\alpha/2$,
resulting in the probability of no communication to be 
$\alpha_{1,0}=1-\alpha$; these probabilities are indicated in
Fig.~\ref{fig:agg}.
Group~2 also has two links, so let $\widetilde{\Vcal}_{2,1}=\{1\}$ and
$\widetilde{\Vcal}_{2,2}=\{3\}$. Similarly to the case above,
let the update probabilities be $\alpha_{2,0}=1-\alpha$, 
$\alpha_{2,1}=\alpha/3$, and $\alpha_{2,2}=2\alpha/3$.
Finally, the only link of group~3 is to group~2, which forms
the group $\widetilde{\Vcal}_{3,1}=\{2\}$. We can take 
$\alpha_{3,0}=1-\alpha$ and $\alpha_{3,1}=\alpha$.

The distributed link matrices $\Phi_{q_1,q_2,q_3}$
in \eqref{eqn:A_agg1} can be expressed as
\[
  \Phi_{q_1,q_2,q_3}
   = \begin{bmatrix}
       \widetilde{\phi}_1^{(q_1)} & \widetilde{\phi}_2^{(q_2)} &
        \widetilde{\phi}_3^{(q_3)}
     \end{bmatrix},~~~q_1,q_2\in\{0,1,2\},~q_3\in\{0,1\},
\]
where each column is given as 
\begin{align*}
  \widetilde{\phi}_1^{(0)} 
    &= \begin{bmatrix}
        1\\ 0\\ 0
      \end{bmatrix},~
  \widetilde{\phi}_1^{(1)} 
    = \begin{bmatrix}
        1/2\\ 1/2\\ 0
      \end{bmatrix},~
  \widetilde{\phi}_1^{(2)} 
    = \begin{bmatrix}
        1/2\\ 0\\ 1/2
      \end{bmatrix},\\
  \widetilde{\phi}_2^{(0)} 
   &= \begin{bmatrix}
        0\\ 1\\ 0
      \end{bmatrix},~
  \widetilde{\phi}_2^{(1)} 
    = \begin{bmatrix}
        1 \\ 0\\ 0
      \end{bmatrix},~
  \widetilde{\phi}_2^{(2)}  
    = \begin{bmatrix}
        0\\ 0\\ 1
      \end{bmatrix},\\      
  \widetilde{\phi}_3^{(0)} 
    &= \begin{bmatrix}
        0\\ 0\\ 1
      \end{bmatrix},~
  \widetilde{\phi}_3^{(1)} 
    = \begin{bmatrix}
        0\\ 1/9\\ 8/9
      \end{bmatrix}.
\end{align*}
It is straightforward to check that 
the properties shown in Proposition~\ref{prop:Aq} hold:
Each $\Phi_{q_1,q_2,q_3}$ is a stochastic matrix and 
the average matrix $\overline{\Phi}$ satisfies the relation 
$\overline{\Phi}=\alpha \Phi + (1-\alpha) I$.
\End
\end{example}

\begin{figure}[t]
\begin{minipage}[t]{8cm}
  \centering
  \fig{8.5cm}{6.5cm}{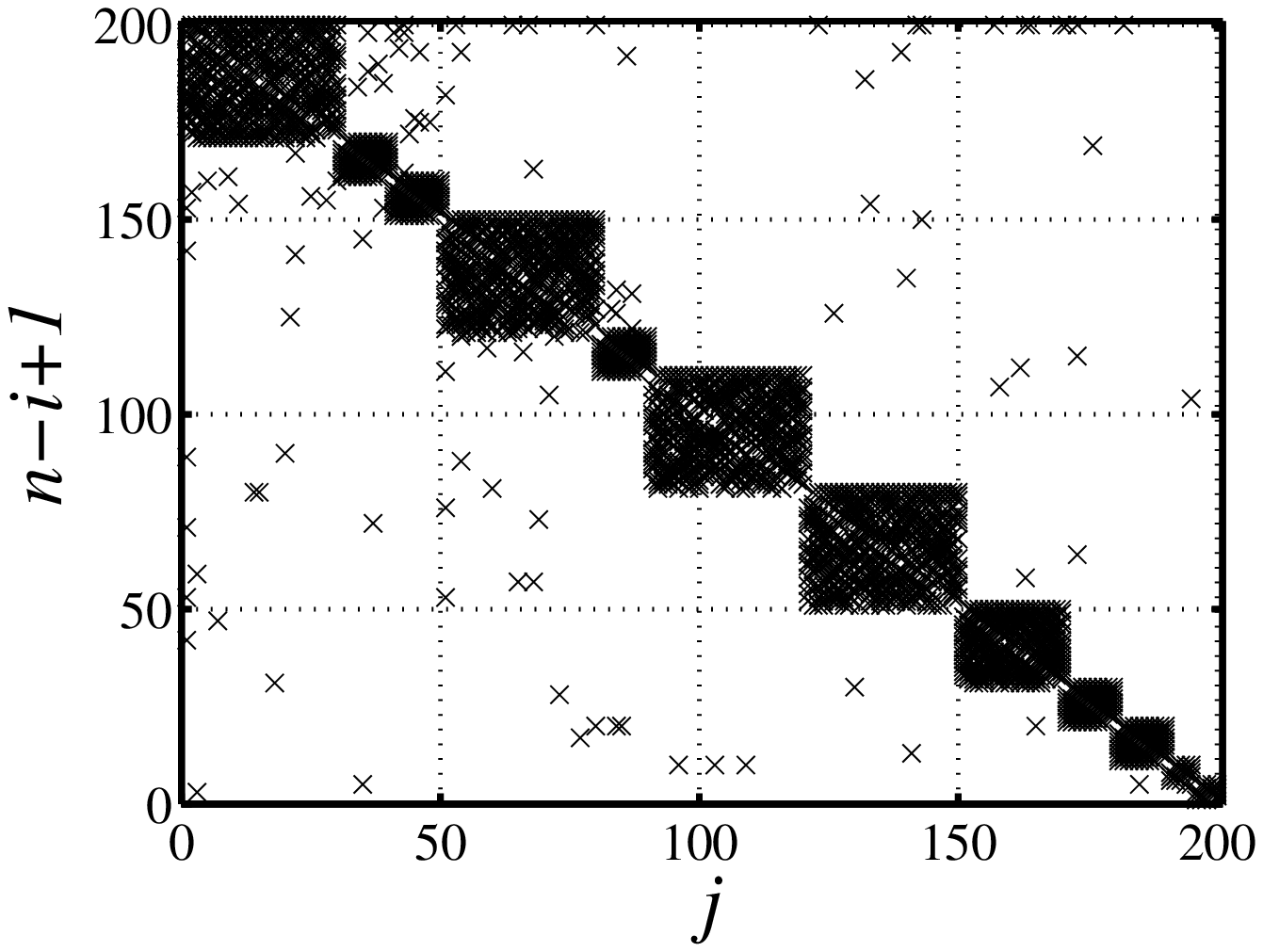}
  \vspace*{-4mm}
  \caption{Connectivity matrix with a block diagonal structure}
  \label{fig:MatrixG}
\end{minipage}\hspace*{4mm}
\begin{minipage}[t]{8cm}
  \centering
  \fig{8.5cm}{6.5cm}{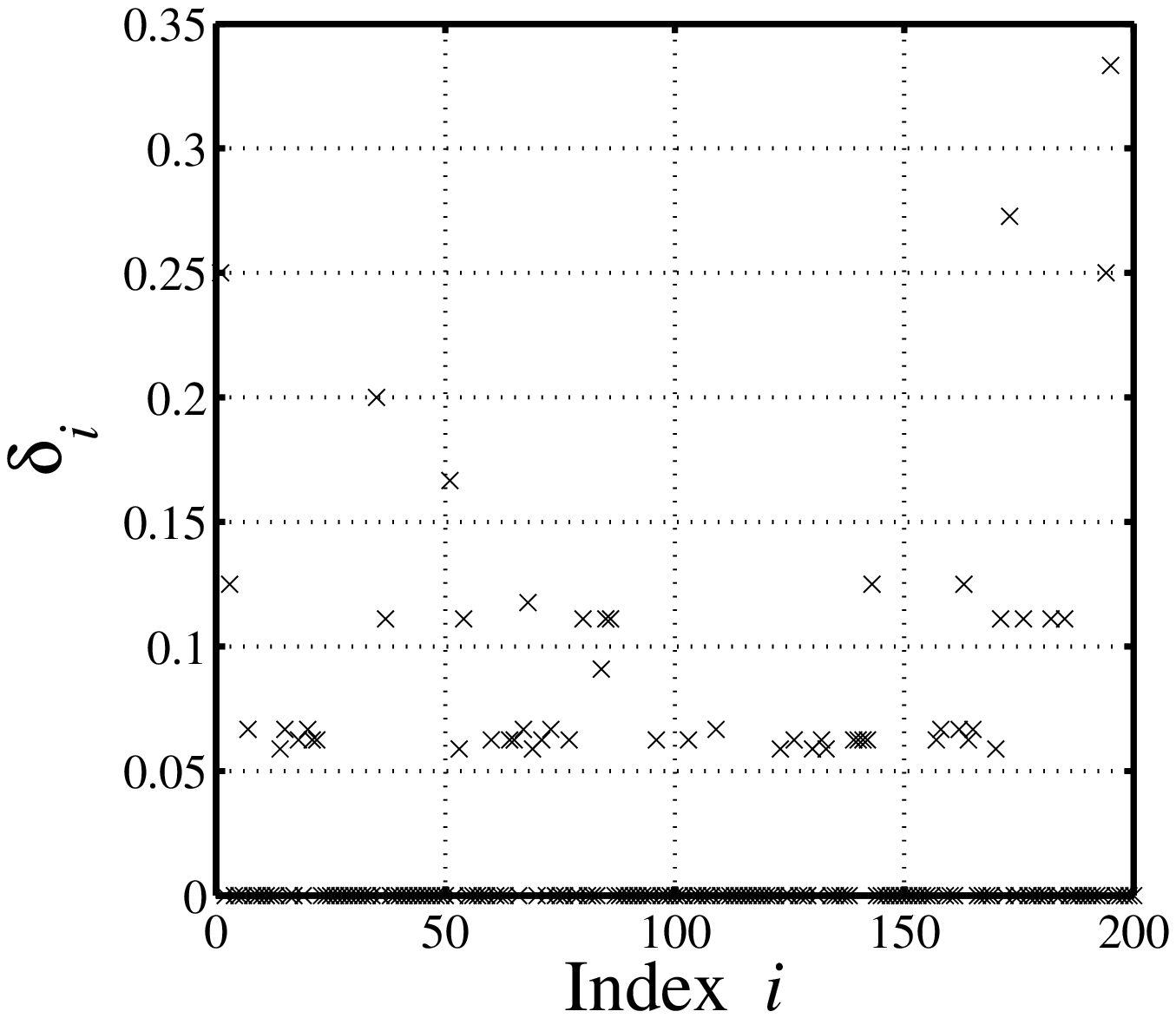}
  \vspace*{-4mm}
  \caption{The original node parameters $\delta_i$ of the pages}
  \label{fig:Delta_i}
\end{minipage}
\end{figure}

%

\section{Numerical example}
\label{sec:example}

In this section, we provide a numerical example
to illustrate the aggregation-based approach.

We consider a web with 200 pages ($n=200$) 
whose link structure was randomly generated with
some level of sparsity.
The connectivity matrix representing the links
is shown in Fig.~\ref{fig:MatrixG}, where
the mark $\times$ indicates the presence of a link
from page $j$ to $i$.  Here, we made
12 dense diagonal blocks, 
each of which consisting of 5 to 30 pages.
Further, there are two less dense blocks of 
pages 1 to 50 and pages 51 to 90. 
These blocks are initially considered as groups. 
The first page in each dense block
has links from all other pages in the same block, and
in particular, pages 1 and 51 are given more incoming and/or 
outgoing links; these pages will have high PageRank values
as we will see. 

Following the grouping procedure outlined in Remark~\ref{rem:group},
we first computed the node parameters $\delta_i$ 
for each page $i$ with respect to the groups described above;
these are plotted in Fig.~\ref{fig:Delta_i}. 
This parameter can be large when the corresponding page has
many outgoing links as pages 1 and 51.
It can also be large when a page is in a small group, but has
external links; this is the case with pages 173, 194, and 195.
However, many pages have no external links, and hence
the average value of $\delta_i$ is relatively small at 0.0260.
We remark that this average is similar to the value 
reported in \cite{BroLem_infret:06}, which is
found from real web data when pages are grouped
according to the hosts. 

\begin{figure}[t]
\begin{minipage}[t]{8cm}
  \centering
  \fig{8.5cm}{6.5cm}{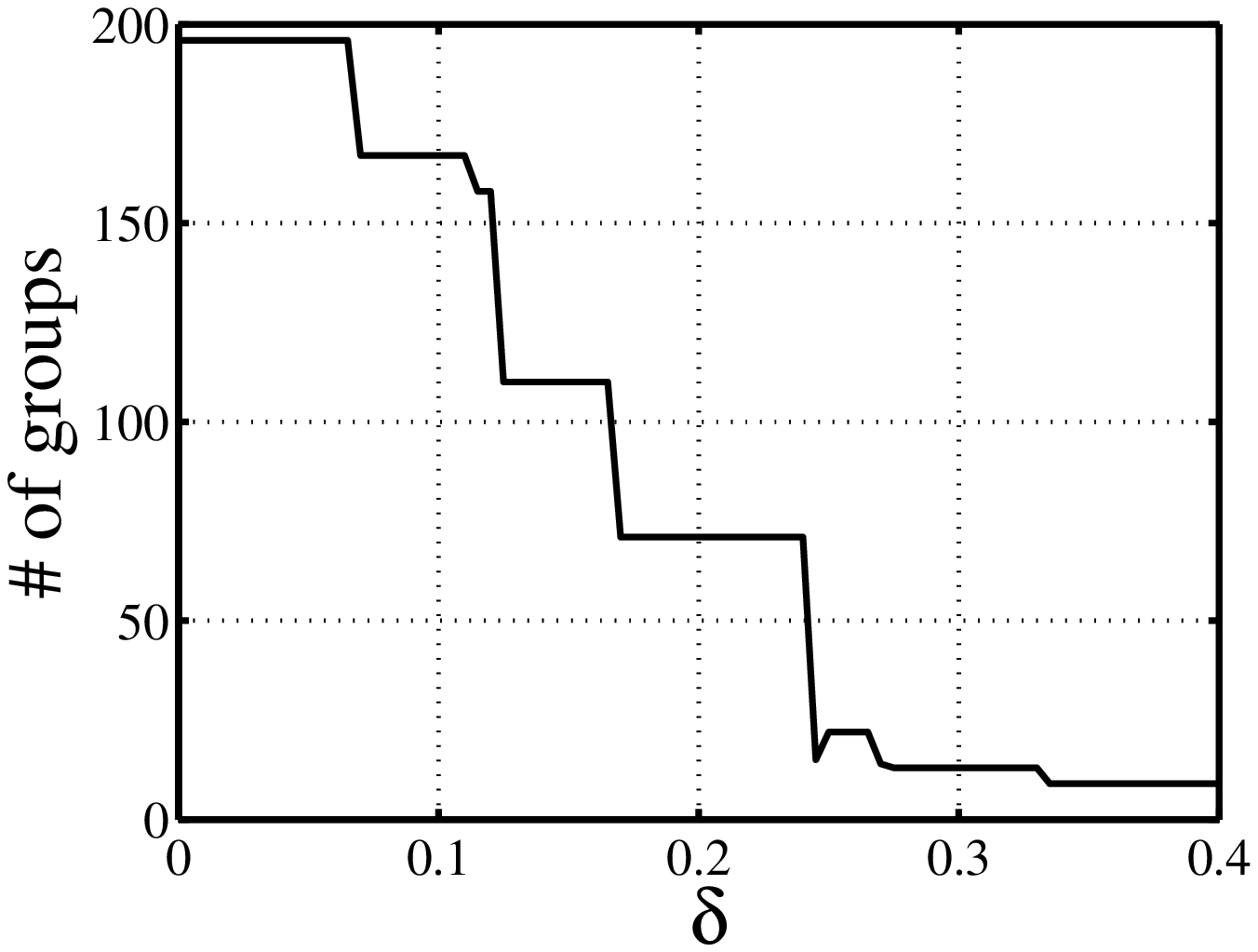}
  \vspace*{-4mm}
  \caption{The number $r$ of groups versus $\delta$}
  \label{fig:NumG_delta}
\end{minipage}\hspace*{4mm}
\begin{minipage}[t]{8cm}
  \centering
  \fig{8.5cm}{6.5cm}{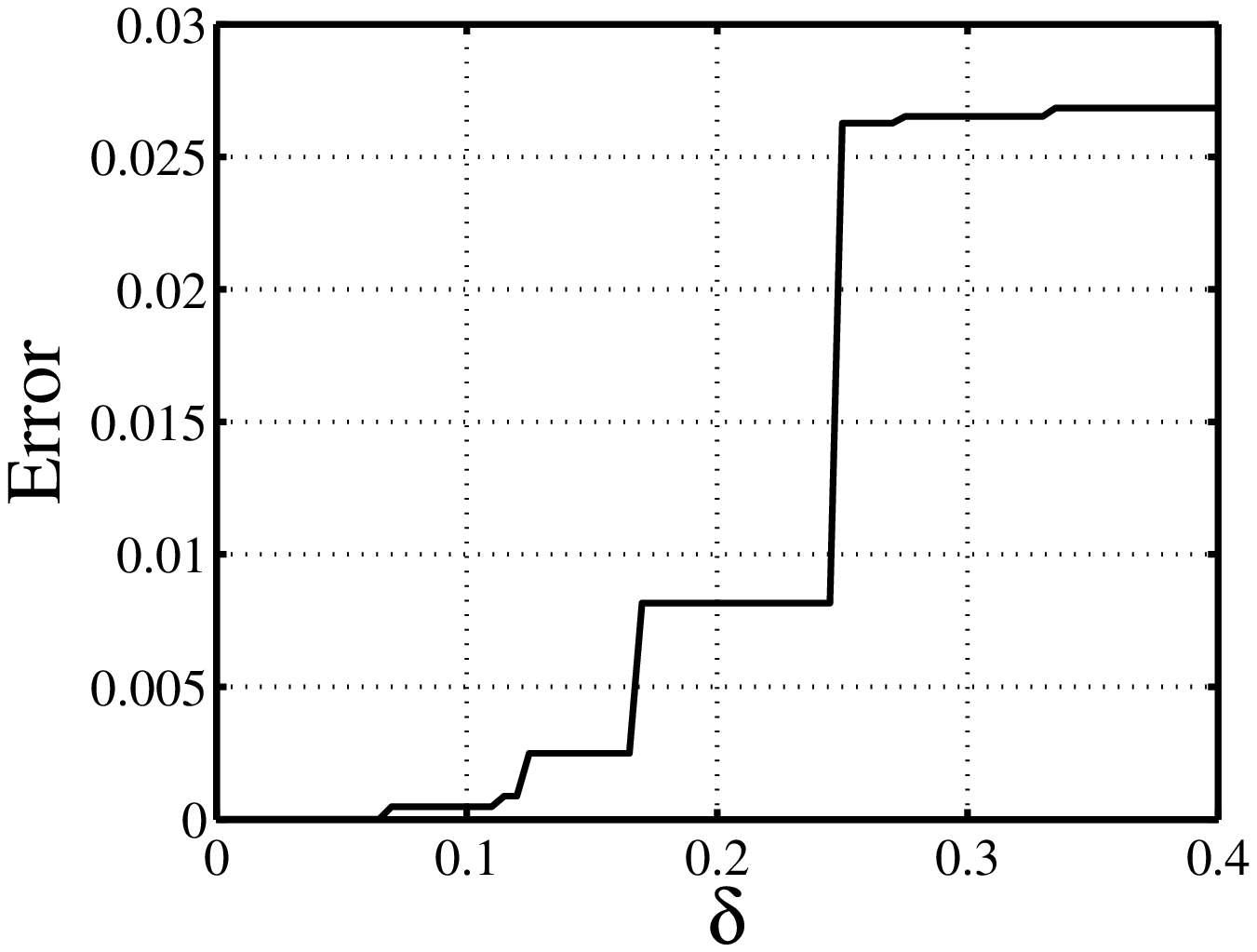}
  \vspace*{-4mm}
  \caption{Error $\norm{x'-x^*}_1$ in PageRank versus $\delta$}
  \label{fig:Error_delta}
\end{minipage}
\end{figure}

%

The grouping procedure is determined by 
the bound $\delta$ on node parameters.
The relation between $\delta$ and the number $r$ of groups is
shown in Fig.~\ref{fig:NumG_delta}.
The line in the figure is not necessarily 
nonincreasing and is in fact piecewise constant.
On the other hand, Fig.~\ref{fig:Error_delta} exhibits
the relation between $\delta$ and 
the error in the approximated PageRank measured by
$\norm{x'-x^*}_1$. These two figures clearly show 
the tradeoff between accuracy in computation and
the size of the problem: Smaller $\delta$ implies
smaller error but larger number $r$ of groups.

\begin{figure}[t]
  \vspace*{3mm}
  \centering
  \fig{8.5cm}{6.5cm}{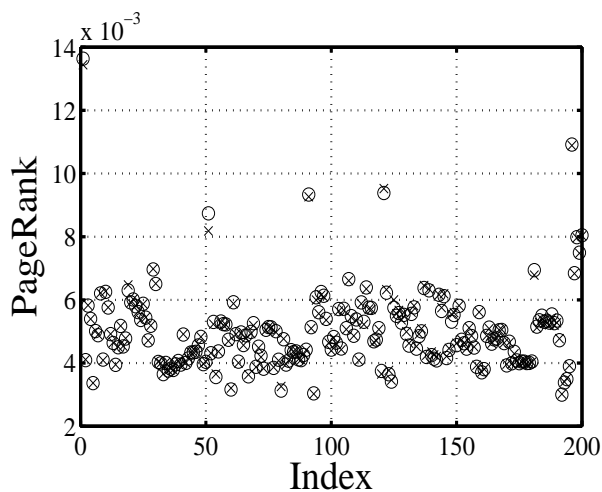}
  \vspace*{-2mm}
  \caption{PageRank: True values $x_i^*$ (in $\times$) and 
              approximation $x_i'$ with $\delta=0.2$ (in $\bigcirc$) 
              for $i=1,\ldots,n$}
  \label{fig:PageRank_delta}
\end{figure}


To see the errors in the PageRank of the individual 
pages, we plotted in Fig.~\ref{fig:PageRank_delta}
the true values $x_i^*$ and the approximated values $x'_i$ for $i\in\Vcal$
when $\delta=0.2$ is used by $\times$ and $\bigcirc$,
respectively. This computation was done with 71 groups ($r=71$).
We observe that the error is small in general.

\begin{figure}[t]
\begin{minipage}[t]{8cm}
  \centering
  \fig{8.5cm}{6.5cm}{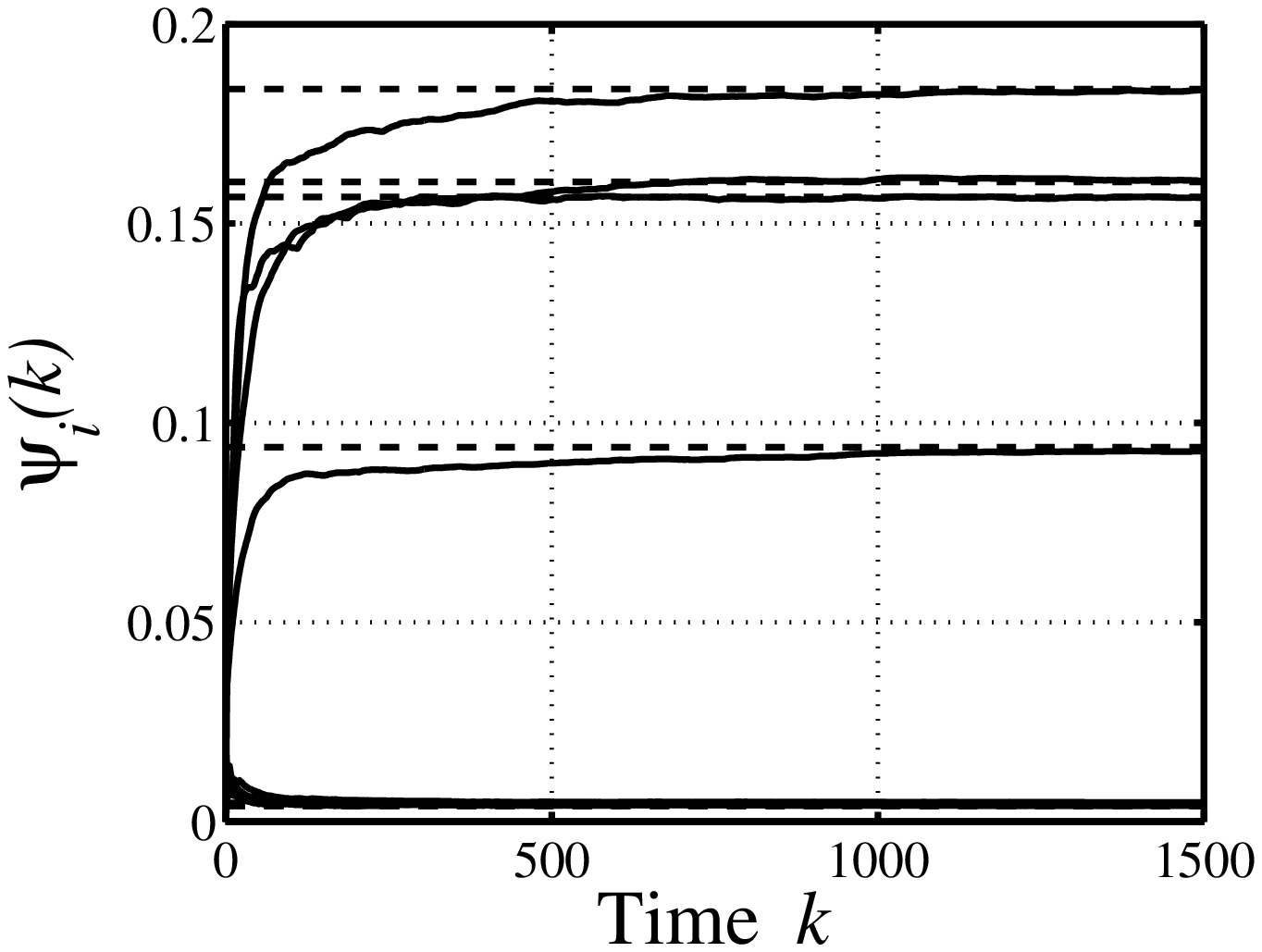}
  \vspace*{-4mm}
  \caption{Distributed randomized algorithm
  for aggregated PageRank: Sample paths of $\psi_i(k)$ 
  with $\delta=0.2$ (solid lines)
  and the true values $\xi_i'$ (dashed lines) for $i=50,\ldots,59$}
  \label{fig:Timeresp_psi}
\end{minipage}\hspace*{4mm}
\begin{minipage}[t]{8cm}
  \centering
  \fig{8.5cm}{6.5cm}{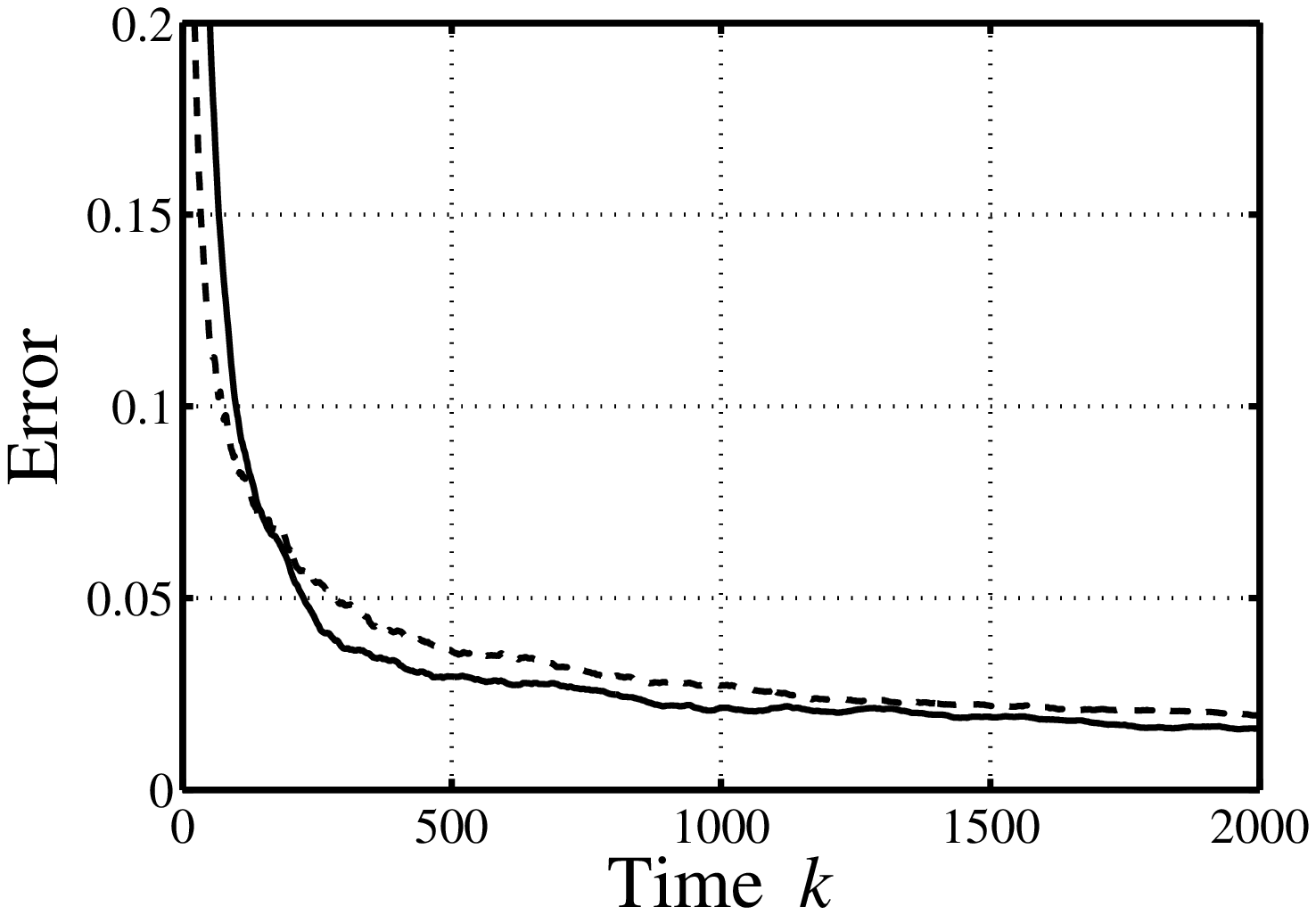}
  \vspace*{-4mm}
  \caption{Comparison of distributed randomized algorithms:
  Sample paths of the error $\norm{x(k)-x^*}_1$ of the aggregated scheme 
  with $\delta=0.2$ (solid line) and 
  the full-order scheme with $n=200$ (dashed line)}
  \label{fig:Timeresp_err}
\end{minipage}
\end{figure}

Finally, we applied the distributed randomized algorithm
\eqref{eqn:xk2} and \eqref{eqn:xk2_psi}
for computing the aggregated PageRank $\xi'$ and then the entire
PageRank vector $x$.
Here, for the gossip communication protocol, 
the grouping $\widetilde{\Vcal}_{i,\ell}$, $i=1,\ldots,r$,
$\ell=0,1,\ldots,g_i$, of the neighbors is based on the
original groups in the block structure of the connectivity matrices.
Furthermore, we set the update probabilities $\alpha_{i,\ell}$ using 
the formula in \eqref{eqn:alpha_il3} with the base probability $\alpha=0.5$.
In Fig.~\ref{fig:Timeresp_psi}, 
sample paths of the time average $\psi_i(k)$ for groups 50 to 59
are shown in solid lines and the true values $\xi_i'$ of 
the aggregated PageRank in dashed lines.
It is clear that the time average converges to the true value.

We would like to emphasize
that the convergence performance of this 
reduced-order scheme is similar to the case without aggregation. 
To make a fair comparison, we computed the state $x(k)$ obtained from 
the time average $\psi_i(k)$ and then its overall error 
from the true PageRank $x^*$. 
We plotted in Fig.~\ref{fig:Timeresp_err} 
the 1-norm of $x(k)-x^*$ by the solid line.
Then, the same error was computed from the time average of 
the full-order scheme, which can be obtained by setting the node
parameter very small as $\delta=0.01$;
the result is shown by the dashed line in the same plot. 
The convergence rates as well as the achieved error levels at the final time 
are comparable for the two cases. The reduced-order case is clearly
advantageous since it requires less operation 
as we have discussed in Remark~\ref{rem:comp}.
More concretely, in the example considered here, 
the numbers of nonzero entries for the link matrices 
$A\in\R^{200\times 200}$ and $\widetilde{A}_{11}=\Phi\in\R^{71\times 71}$ 
are, respectively, $f_0(A)=2623$ and $f_0(\widetilde{A}_{11})=780$.
In view of Table~\ref{table:operation}, the difference in 
computation costs is evident.

\section{Conclusion}
\label{sec:concl}

In this paper, we have developed a distributed randomized algorithm
for obtaining the PageRank values which performs well with reduced 
computation and communication loads. 
The approach is based on a novel aggregation technique of the web.
First, we have proposed a simple procedure for grouping pages under 
the criterion of node parameters.  Then, the notion of aggregated PageRank 
has been introduced, based on which approximates of the true values
for individual pages can be computed.  Error bounds on the approximation
level have been derived. Moreover, we have developed 
the distributed randomized algorithm
of lower order for the computation of aggregated PageRank. 
The advantage of the approach in terms of computation cost as
well as convergence properties have been discussed in detail and also 
demonstrated by a numerical example. 

In the future, we will further study aggregation-based methods 
for PageRank to improve the convergence rate of 
the update scheme and the effects of incorporating multi-level groupings.
More generally, for large-scale systems, the concept
of aggregation is of great importance and may provide useful tools
for other interesting problems such as smart grid for power distribution. 

\smallskip
\noindent
\textit{Acknowledgement}:~The authors would like to thank
Athanasios C.\ Antoulas, Fabrizio Dabbene, Soura Dasgupta, 
Shinji Hara, and Jun-ichi Imura for helpful discussions. 


\small

\end{document}